\documentclass[aps,prd,nofootinbib,showpacs,superscriptaddress, preprint]{revtex4-1}
\pdfoutput=1

\newcommand{\bmt}{\begin{pmatrix}}
\newcommand{\emt}{\end{pmatrix}}
\newcommand{\ba}{\begin{array}{c}}
\newcommand{\ea}{\end{array}}
\newcommand{\be}{\begin{equation}}
\newcommand{\ee}{\end{equation}}
\newcommand{\bea}{\begin{eqnarray}}
\newcommand{\eea}{\end{eqnarray}}

\newcommand{\bi}{\begin{itemize}}
\newcommand{\ei}{\end{itemize}}

\newcommand{\baz}{\begin{array}{cc}}
\newcommand{\besub}{\begin{subequations}}
\newcommand{\eesub}{\end{subequations}}
\newcommand{\mathsym}[1]{{}}

\newcommand{\bt}{\begin{tabular}}
\newcommand{\et}{\end{tabular}}

\newcommand{\benu}{\begin{enumerate}}
\newcommand{\eenu}{\end{enumerate}}
\newcommand{\ogw}{\Omega_\text{gw}}

\def\q2 {q^2}

\def\bt{\begin{table}}
\def\et{\end{table}}

\usepackage[utf8]{inputenc}
\usepackage{amsmath,amssymb}
\usepackage{epsfig}
\usepackage{comment}
\usepackage{graphicx}
\usepackage[usenames,dvipsnames]{color}
\usepackage{float}
\usepackage{bbold}
\usepackage[colorlinks,citecolor=blue]{hyperref}
\usepackage{color}
\usepackage{subfigure}
\begin{document}
%opening
%\title{Probing PBH generated DM via GW observations with multiple spectral breaks}

\title{Probing high scale seesaw and PBH generated dark matter via gravitational waves with multiple tilts}

%%%%%%%%%   Authors   %%%%%%%%%%%%

%\author{Basabendu Barman}
%\email{basabendu88barman@gmail.com}
%\affiliation{\,Centro de Investigaciones, Universidad Antonio Nari\~{n}o\\Carrera 3 este \# 47A-15, Bogot{\'a}, Colombia}

\author{Debasish Borah}
\email{dborah@iitg.ac.in}
\affiliation{Department of Physics, Indian Institute of Technology Guwahati, Assam 781039, India}

\author{Suruj Jyoti Das}
\email{suruj@iitg.ac.in}
\affiliation{Department of Physics, Indian Institute of Technology Guwahati, Assam 781039, India}

\author{Rishav Roshan}
\email{rishav.roshan@gmail.com}
\affiliation{Department of Physics, Kyungpook National University, Daegu 41566, Korea}

\begin{abstract}
 We propose a scenario where a high scale seesaw origin of light neutrino mass and gravitational dark matter (DM) in MeV-TeV ballpark originating from primordial black hole (PBH) evaporation can be simultaneously probed by future observations of stochastic gravitational wave (GW) background with multiple tilts or spectral breaks. A high scale breaking of an Abelian gauge symmetry ensures the dynamical origin of the seesaw scale while also leading to the formation of cosmic strings responsible for generating stochastic GW background. The requirement of a correct DM relic in this ballpark necessitates the inclusion of a diluter as PBH typically leads to DM overproduction. This leads to a second early matter dominated epoch after PBH evaporation due to the long-lived diluter. These two early matter dominated epochs, crucially connected to the DM relic, lead to multiple spectral breaks in the otherwise scale-invariant GW spectrum formed by cosmic strings. We find interesting correlations between DM mass and turning point frequencies of GW spectrum which are within reach of several near future experiments like LISA, BBO, ET, CE, etc.
\end{abstract}

\maketitle
%\flushbottom

\section{Introduction}
The observation of neutrino oscillation has been the most compelling experimental evidence suggesting the presence of beyond standard model (BSM) physics. This is because the standard model (SM) alone can not explain the origin of neutrino mass and mixing, as verified at neutrino oscillation experiments \cite{Zyla:2020zbs, Mohapatra:2005wg}. Different BSM frameworks have, therefore, been invoked to explain non-zero neutrino mass.
%as in the SM, there is no way to couple the left handed neutrinos to the Higgs field in the renormalisable Lagrangian due to the absence of right chiral neutrinos. 
Conventional neutrino mass models based on seesaw mechanism \cite{Minkowski:1977sc, GellMann:1980vs, Mohapatra:1979ia, Sawada:1979dis,Yanagida:1980xy, Schechter:1980gr, Mohapatra:1980yp, Lazarides:1980nt, Wetterich:1981bx, Schechter:1981cv, Foot:1988aq} typically involve the introduction of heavy fields like right handed neutrinos (RHN). Typically, such canonical seesaw models have a very high seesaw scale keeping it away from the reach of any direct experimental test. This has led to some recent attempts in finding ways to probe high scale seesaw via stochastic gravitational wave (GW) observations \cite{Dror:2019syi, Blasi:2020wpy, Fornal:2020esl, Samanta:2020cdk, Barman:2022yos, Huang:2022vkf, Dasgupta:2022isg, Okada:2018xdh, Hasegawa:2019amx, Borah:2022cdx}. In these scenarios, the presence of additional symmetries was considered whose spontaneous breaking not only leads to the dynamical generation of RHN masses but also leads to the generation of stochastic GW from topological defects like cosmic strings \cite{Dror:2019syi, Blasi:2020wpy, Fornal:2020esl, Samanta:2020cdk}, domain walls \cite{Barman:2022yos} or from bubbles generated at first order phase transition \cite{Huang:2022vkf, Dasgupta:2022isg, Okada:2018xdh, Hasegawa:2019amx, Borah:2022cdx}.

In addition to the origin of neutrino mass, the SM also fails to explain the origin of dark matter (DM), a mysterious, non-luminous and non-baryonic form of matter comprising around $27\%$ of the present universe. The present DM abundance is often reported in terms of density parameter $\Omega_{\rm DM}$ and reduced Hubble constant $h = \text{Hubble Parameter}/(100 \;\text{km} ~\text{s}^{-1} 
\text{Mpc}^{-1})$ as \cite{Aghanim:2018eyx}
\begin{equation}
\Omega_{\text{DM}} h^2 = 0.120\pm 0.001
\label{dm_relic}
\end{equation}
\noindent at 68\% CL. Among a variety of BSM proposals to explain the origin of DM, the weakly interacting massive particle (WIMP) has been the most well-studied one. In the WIMP framework, a particle having mass and interactions around the electroweak ballpark can lead to the observed DM abundance after undergoing thermal freeze-out in the early universe \cite{Kolb:1990vq}. While WIMP has promising direct detection signatures, contrary to high scale seesaw, the null results at direct detection experiments \cite{LUX-ZEPLIN:2022qhg} have motivated us to consider other alternatives. Such alternative DM scenarios may have suppressed direct detection signatures but can leave signatures in future GW experiments. There have been several recent works on such GW signatures of DM scenarios \cite{Yuan:2021ebu, Tsukada:2020lgt, Chatrchyan:2020pzh, Bian:2021vmi, Samanta:2021mdm, Borah:2022byb, Azatov:2021ifm, Azatov:2022tii, Baldes:2022oev,Chakrabortty:2020otp}.

    Motivated by this, we consider a high scale seesaw scenario with gravitational DM generated from evaporating primordial black holes (PBH) assuming the latter to dominate the energy density of the universe at some stage. While the seesaw scale is dynamically generated due to the spontaneous breaking of a $U(1)$ gauge symmetry at a high scale, DM can be generated during a PBH-dominated epoch. Unlike sub-GeV or superheavy DM (with mass $\geq \mathcal{O}(10^9 \, \rm GeV)$) considered in most of the PBH-generated DM scenarios like \cite{Lennon:2017tqq, Morrison:2018xla, Gondolo:2020uqv, Bernal:2020bjf, Bernal:2021bbv, Bernal:2021yyb, Hooper:2019gtx}, we consider DM around MeV-TeV ballpark \cite{Barman:2022gjo}. Since DM in this ballpark gets overproduced during a PBH-dominated era, we also consider the presence of a long-lived diluter whose late decay brings the DM abundance within limits. The diluter is also dominantly produced from PBH as its coupling with the SM remains suppressed in order to keep it long-lived. The high-scale breaking of $U(1)$ gauge symmetry leads to the formation of cosmic strings (CS) \cite{Kibble:1976sj, Nielsen:1973cs}, which generate stochastic GW background with a characteristic spectrum within the reach of near future GW detectors if the scale of symmetry breaking is sufficiently high \cite{Vilenkin:1981bx,Turok:1984cn}. In the present setup, we show that the PBH domination followed by the subsequent diluter domination era lead to multiple spectral breaks in the GW spectrum formed by CS network. In a recent work, \cite{Borah:2022byb} the correlation between such spectral break due to non-standard cosmological era \cite{Cui:2017ufi,Cui:2018rwi,Gouttenoire:2019kij} of diluter domination and DM properties were studied within the framework of a realistic particle physics model with high scale seesaw. Here we extend this to a PBH-generated gravitational DM scenario with the unique feature of multiple spectral breaks in scale-invariant GW background generated by CS, not considered in earlier works. Depending upon DM mass, the three spectral breaks or turning point frequencies can be within reach of different future GW experiments. The novel feature of our study is in finding the role of multiple early matter domination eras in the otherwise scale-invariant GW spectrum of CS network leading to multiple turning point frequencies and the connection of these eras as well as the turning point frequencies of GW spectrum with mass of gravitational DM produced from PBH evaporation.

This paper is organized as follows. In section \ref{sec1}, we briefly discuss our setup followed by a discussion of dark matter relic from PBH in section \ref{sec2}. In section \ref{sec3}, we discuss the details of the gravitational wave spectrum generated by cosmic strings and subsequent spectral breaks due to PBH and diluter domination for different DM masses. In section \ref{sec4}, we discuss briefly the gravitational waves generated from PBH density fluctuations. Finally, we conclude in section \ref{sec5}.

%%%%%%%%%%%%%%%%%%%%%%%%%%%%%%%%%%%%%%%%%%%%%%%%%%%%%%%%%%%%%%%%%%%%%%%%%%%%%%%%%%%%%%%%%%%%%%%%%%%%%%%%%%%%%%%%%%%%%%%%%555555
\section{The Framework}
\label{sec1}

The present framework can be embedded in any Abelian extension of the SM that guarantees the formation of cosmic strings due to the spontaneous symmetry breaking with an era of primordial black hole domination. A singlet scalar with no coupling to the SM or to the new Abelian gauge sector is assumed to be the DM having purely gravitational interactions. A diluter in the form of a heavy singlet Majorana fermion or a heavy RHN is also incorporated while keeping its coupling to the SM leptons tiny. In order to ensure that the PBH domination and diluter domination arise as separate epochs for simplicity, we consider the diluter to be singlet under the new Abelian gauge symmetry as well.

Here, we outline the minimal interaction terms to realise our scenario with one possible UV completion being given in Appendix \ref{sec:model}. The two RHNs having sizeable couplings to leptons thereby generating light neutrino masses via type I seesaw mechanism\footnote{We consider a normal hierarchy for the light neutrino masses.} \cite{Minkowski:1977sc, GellMann:1980vs, Mohapatra:1979ia, Yanagida:1980xy, Schechter:1980gr, Schechter:1981cv} are denoted by $N_{1,2}$ while the one playing the role of diluter and having tiny Yukawa coupling is denoted by $N_3$. Dark matter is assumed to be a real singlet scalar with only gravitational interactions. The RHNs $N_{1,2}$ are charged under a $U(1)$ gauge symmetry such that the spontaneous $U(1)$ symmetry breaking by a singlet scalar $S$ dynamically generates the seesaw scale. The relevant Lagrangian of the newly introduced fields is given by 
\begin{align}
-\mathcal{L} \supset \sum_{\alpha, i} Y_{\alpha i} \overline{L}_\alpha \tilde{H} N_i + \frac{1}{2} \sum_{i,j=1,2} h_{ij} S\overline{N^c_i}N_j+\frac{1}{2} M_3 \overline{N^c_3} N_3 + \frac{m^2_{\rm DM}}{2} \phi^2.
\end{align}
In order to ensure $N_3$ to be long-lived $Y_{\alpha 3} \ll 1$. Due to tiny coupling with the bath particles, the diluter is dominantly produced only from PBH evaporation. Similarly, due to the absence of any DM coupling with the bath, it will also be solely produced from PBH. Additionally, the singlet scalar $S$ can acquire a non-zero vacuum expectation value (VEV), leading to the generation of $N_{1,2}$ as well as $U(1)$ gauge boson masses. The same symmetry breaking also leads to the formation of cosmic strings. We discuss the details of DM relic generation and cosmic string dynamics leading to gravitational waves in the upcoming sections.

\section{Dark matter and diluter from PBH}
\label{sec2}
As mentioned above, the DM in the present framework only has gravitational interactions and hence can be dominantly produced from PBH evaporation. On the other hand, as a result of its feeble interaction with the SM bath, $N_3$ can also be assumed to be produced purely from the PBH evaporation. While the PBH itself can act as a DM if its mass is chosen appropriately, in this work we focus on the ultra-light mass regime of PBH where it is not cosmologically long-lived. As is well known from several earlier works, the DM in a mass range: a few keV to around $10^{10}$ GeV, when produced from the PBH evaporation are always overabundant if PBHs dominate the total energy density of the Universe~\cite{Fujita:2014hha, Samanta:2021mdm}. A recent study \cite{Barman:2022gjo} has shown that the overproduction of such DM can be prevented if one considers a diluter that injects entropy at a late time into the thermal bath as a result of its slow decay. Following \cite{Barman:2022gjo}, in the present work we also consider such multiple early matter-dominated eras and show how their presence can affect the GW spectrum generated by cosmic strings in a unique way.

PBH is assumed to be formed in the era of radiation domination after inflation. In such a situation, the PBH abundance is characterized by a dimensionless parameter $\beta$ defined as 
\bea
\beta=\frac{\rho_\text{BH}(T_\text{in})}{\rho_\text{R}(T_\text{in})},
\eea
where $\rho_\text{BH}(T_\text{in})$ and $\rho_\text{R}(T_\text{in})$ represent the initial PBH energy density and radiation energy density respectively while $T_\text{in}$ denotes the temperature at the time of PBH formation. The expression of $T_\text{in}$ can be found in appendix~\ref{sec:pbh}. Once PBHs are formed, the radiation-dominated universe eventually turns into a matter-dominated universe and remains matter-dominated till the epoch of PBH evaporation. The temperature corresponding to the PBH evaporation is denoted as $T_\text{ev}$ and can be obtained from Eq.~\eqref{eq:pbh-Tev}. The condition determining PBH's evaporation during the radiation domination is given as~\cite{Masina:2020xhk}, 
\begin{equation}
\beta<\beta_\text{c}\equiv \gamma^{-1/2}\,\sqrt{\frac{\mathcal{G}\,g_{\star,H}(T_\text{BH}^{\rm in})}{10640\,\pi}}\,\frac{M_\text{pl}}{m_\text{in}}\,,
\label{eq:pbh-ev-rad}
\end{equation}
where $M_{\rm pl}$ indicates the Planck mass and
\begin{equation}
{g_{\star,H}}(T_\text{BH})\equiv\sum_i\omega_i\,g_{i,H} e^{-\frac{M_{BH}}{\beta_i M_i}}\,; g_{i,H}=
    \begin{cases}
        1.82
        &\text{for }s=0\,,\\
        1.0
        &\text{for }s=1/2\,,\\
        0.41
        &\text{for }s=1\,,\\
        0.05
        &\text{for }s=2\,,\\
    \end{cases}
    \label{eqn:gsh}
\end{equation}
with $\omega_i=2\,s_i+1$ for massive particles of spin $s_i$, $\omega_i=2$ for massless species with $s_i>0$ and $\omega_i=1$ for $s_i=0$. $\beta_s$ = 2.66, 4.53, 6.04, 9.56  for $s_i$ = 0,1/2,1,2 respectively, whereas $M_{i}= \frac{1}{8\pi G m_{i}}$, where $m_{i}$ is the mass of the $i^\text{th}$ species \cite{Masina:2020xhk}. $\beta_c$ is the critical PBH abundance that leads to an early matter-dominated era. In our scenario, we consider $\beta$ to be large enough such that PBH dominates the energy density of the universe at some epoch. In the above inequality, $\gamma\simeq0.2$ is a numerical factor that contains the uncertainty of the PBH formation, and $\mathcal{G}\sim 4$ is the grey-body factor. Here, $M_\text{pl}$ denotes the Planck mass while $m_\text{in}$ and $T_\text{BH}$ denote the mass of PBH at the time of formation and instantaneous Hawking temperature of PBH (see appendix~\ref{sec:pbh} for the details).  Additionally, we also assume the PBHs to be of Schwarzschild type without any spin and charge and having a monochromatic mass spectrum implying all PBH have identical masses. It is important to mention that a PBH can dominantly evaporate only into particles lighter than its instantaneous Hawking temperature.

An upper bound on the PBH mass is obtained by demanding its evaporation temperature $T_\text{ev}> T_\text{BBN}\simeq 4$ MeV, as the evaporation of PBH also results in the production of the radiation that can disturb the successful prediction of the big bang nucleosynthesis (BBN). While the upper bound on the PBH mass can be obtained by comparing its evaporation temperature with the BBN temperature, a lower bound on its mass can be obtained from the cosmic microwave background (CMB) bound on the scale of inflation, $i.e.$ $\mathcal{H}(T_\text{in}) \leq \mathcal{H}_I\leq 2.5\times 10^{-5}\,M_\text{pl}$, where $\mathcal{H}(T_\text{in})=\frac{1}{2\,t_\text{in}}$ with $t(T_\text{in})=\frac{m_\text{in}}{M_\text{pl}^2\,\gamma}$ (as obtained from Eq.~\eqref{eq:pbh-mass}). Using these BBN and CMB bounds together, we have a window for allowed initial mass for PBH that reads $0.1\,\text{g}\lesssim m_\text{in}\lesssim 3.4\times 10^8\,\text{g}$. The range of PBH masses between these bounds is unconstrained from  BBN, CMB and other such bounds at low redshifts ~\cite{Carr:2020gox}. However, Refs. \cite{Domenech:2020ssp, Domenech:2021wkk} have recently put constraints on such low PBH mass as well, from the gravitational wave background produced from PBH density fluctuations, since these GW can lead to extra relativistic degrees of freedom during BBN. We discuss this GW spectrum in Section \ref{sec4}.

Since in this work, our primary goal is to investigate the effect of a PBH-dominated Universe on the production of dark matter and its signature on gravitational waves, we remain agnostic about the formation of PBH. However, our choice of the initial energy density of PBH $\beta$, which we consider to be a free parameter satisfying $\beta > \beta_c$ ensures that we have a PBH domination.  Now, the production of PBH with the mass and $\beta$ ranges we have considered can take place, say from perturbations generated during inflation, when they re-enter the horizon, or from phase transitions. For inflationary production mechanisms, the mass of PBH formed can be related to the scale of the perturbation as \cite{Hawking:1971ei, Carr:1974nx, Wang:2019kaf, Byrnes:2021jka, Braglia:2022phb}  
\begin{align}
    \frac{m_{\rm in}(k)}{M_\odot}\simeq10^{-16}\left(\frac{\gamma}{0.2}\right)\left(\frac{g_*(T_k)}{106.75}\right)^{-1/6}\left(\frac{k}{10^{14}\,{\rm Mpc}^{-1}}\right)^{-2},
\end{align}
where $M_\odot$ denotes the solar mass and $g_*(T_k)$ indicates the relativistic degrees of freedom when the scale `$k$' enters the horizon. Hence, for forming PBH with a mass function peaked around  $m_{\rm in }\sim 10^{3}- 10^{6}$ g, which, as we will see, is the range we have used, an enhancement to the power spectrum is needed at extremely small scales $k \sim 10^{19}-10^{21} ~\text{Mpc}^{-1}$, corresponding to the last stages of inflation. We leave inflationary potentials giving rise to such features in the power spectrum for future studies. Now, the initial abundance of PBH $\beta$ is determined by the amplitude of the power spectrum $P_R$ \cite{Press:1973iz, Byrnes:2021jka}, which we can roughly estimate to be given by $P_R\sim \frac{0.2}{\text{ln}(1/\beta)}$ \cite{Byrnes:2021jka}. The amplitude is thus only logarithmically dependent on $\beta$ and we can have our required $\beta > \beta_c$ with  $P_R\sim 10^{-2}$.  

PBH can also be formed during a first order phase transition (FOPT) \cite{Crawford:1982yz, Hawking:1982ga, Moss:1994iq, Kodama:1982sf, Baker:2021nyl, Kawana:2021tde, Huang:2022him}
a result of bubble collisions, Fermi-ball collapse among several others. For example, in the collapsing Fermi-ball scenario \cite{Kawana:2021tde}, a dark sector fermion $\chi$ with an asymmetry between $\chi$ and $\bar{\chi}$ leads to the formation of Fermi-balls after getting trapped in the false vacuum. Due to the strong Yukawa force among the dark fermions, such Fermi-balls can collapse into PBH. The energy fraction for such PBH is given by \cite{Kawana:2021tde}
\begin{equation}
\beta \approx 1.4 \times 10^{-15} v^{-3}_b \left ( \frac{g_*}{100} \right )^{1/2} \left ( \frac{T_*}{100 \, {\rm GeV}} \right )^3 \left ( \frac{\beta_{\rm PT}/\mathcal{H}}{100} \right )^3 \left ( \frac{m_{\rm in}}{10^{15} \, {\rm g}} \right )^{3/2}
\end{equation}
with $T_*, v_b, \beta_{\rm PT}/\mathcal{H}$ being the nucleation temperature, bubble wall velocity and time scale ratio of universe's expansion and the FOPT respectively. For PBH mass of our interest, say $m_{\rm in}= 10^7$ g, we can have $\beta \sim 10^{-9}$ if $T_* = 10^7$ GeV, $v_b \sim 0.1, \beta_{\rm PT}/\mathcal{H} \sim 100$. The initial PBH mass of $10^7$ g can be obtained for the same choice of parameters if dark sector asymmetry is $\sim \mathcal{O}(10^{-4})$.

There are other PBH formation mechanisms discussed in the literature namely, from the collapse of topological defects \cite{Hawking:1987bn, Deng:2016vzb} which we do not elaborate here. Thus, it is possible to generate PBH with the desired initial mass and energy fraction if we incorporate additional physics as mentioned above. In the spirit of minimality, we stick to our minimal setup to highlight the key feature of our proposal namely, the effect of multiple early matter domination on cosmic string generated GW spectrum with a non-trivial connection to dark matter genesis from evaporating PBH.

\subsection{Particle production from PBH}

Once formed, the PBH can evaporate by Hawking radiation~\cite{Hawking:1974rv, Hawking:1975vcx}. Among the products of PBH evaporation, there might be a stable BSM particle that can contribute to the observed DM abundance \cite{Morrison:2018xla, Gondolo:2020uqv, Bernal:2020bjf, Green:1999yh, Khlopov:2004tn, Dai:2009hx, Allahverdi:2017sks, Lennon:2017tqq, Hooper:2019gtx, Chaudhuri:2020wjo, Masina:2020xhk, Baldes:2020nuv, Bernal:2020ili, Bernal:2020kse, Lacki:2010zf, Boucenna:2017ghj, Adamek:2019gns, Carr:2020mqm, Masina:2021zpu, Bernal:2021bbv, Bernal:2021yyb, Samanta:2021mdm, Sandick:2021gew, Cheek:2021cfe, Cheek:2021odj, Barman:2022gjo} or there might be a new particle produced which is responsible for generating the present matter-antimatter asymmetry of the universe \cite{Hawking:1974rv, Carr:1976zz, Baumann:2007yr, Hook:2014mla, Fujita:2014hha, Hamada:2016jnq, Morrison:2018xla, Hooper:2020otu, Perez-Gonzalez:2020vnz, Datta:2020bht, JyotiDas:2021shi, Smyth:2021lkn, Barman:2021ost, Bernal:2022pue, Ambrosone:2021lsx, Barman:2022gjo, Bhaumik:2022pil}. The number of any particle $X$ produced during the evaporation of a single PBH can be estimated by considering the differential mass decrease $d m_{\rm BH} = -M_P^{2}\frac{d T_{\rm BH}}{T_{\rm BH}^2}= -dE\, $\cite{Baumann:2007yr}, where $T_{\rm BH}$ is the Hawking temperature and $dE$ is the corresponding energy emitted.  Considering the mean energy of the emitted particles as $3T_{\rm BH}$, the differential number of emitted particles is found to be
 \begin{equation}
     d \mathcal{N}= \frac{dE}{3T_{\rm BH}}= M_P^{2} \frac{d T_{\rm BH}}{3T_{\rm BH}^3}.
 \end{equation}
 Integrating the above expression gives us the number of particles $X$ as
\begin{equation}
    \mathcal{N}_X = \frac{g_{X,H}}{g_{\star,H}(T_\text{BH})}
    \begin{cases}
       \frac{4\,\pi}{3}\,\Bigl(\frac{m_\text{in}}{M_\text{pl}}\Bigr)^2 &\text{for } m_X < T_\text{BH}^\text{in}\,,\\[8pt]
        \frac{1}{48\,\pi}\,\Bigl(\frac{M_\text{pl}}{m_X}\Bigr)^2 &\text{for } m_X > T_\text{BH}^\text{in}\,,
    \end{cases}\label{eq:pbh-num}\,,
\end{equation}
 Here, $T_\text{BH}^\text{in}$ denotes the instantaneous PBH temperature at the time of the formation. While $g_{\star, H}$ accounts for all emitted particles (Eq. \ref{eqn:gsh}), $g_{X, H}$ corresponds to the particle X. Note that for $m_X < T_\text{BH}^\text{in}$, the emission of particles takes place from the beginning of PBH formation, while in the opposite case of $m_X > T_\text{BH}^\text{in}$, the emission occurs only when $T_{\rm BH}$ reaches $m_{X}$. This leads to different expressions in these two limits after integration. The yield of a particle produced during the evaporation of a PBH is related to the PBH abundance ($n_\text{BH}$) and is expressed as,
\bea
Y_{\mathcal{N}_X}(T_0)=\frac{n_{\mathcal{N}_X}}{s}\bigg{|}_{T_0}=\mathcal{N}_X\frac{n_\text{BH}}{s}\bigg{|}_{T_\text{ev}},
\label{particle_yield}
\eea
where $s$ denotes the entropy density at evaporation. PBH abundance at the time of its evaporation can be obtained using the first Friedmann equation as
\begin{equation}
n_\text{BH} (T_\text{ev}) = \frac{1}{6\,\pi}\,\frac{M_\text{pl}^2}{m_\text{in}\,\tau^2} \equiv\frac{1}{6\,\pi}\,\left(\frac{\mathcal{G}\,g_{\star,H}}{10640\,\pi}\right)^2\,\frac{M_\text{pl}^{10}}{m_\text{in}^7}\,, 
\label{PBH_abundance}
\end{equation}
where $\tau$ is the PBH lifetime (cf. Eq. \eqref{eq:PBHlt}). Substituting Eq.~\eqref{PBH_abundance} in Eq.~\eqref{particle_yield}, one can easily obtain the asymptotic yield of the particles produced during the PBH evaporation.

Following this, one can easily calculate the abundance of the DM and the diluter ($N_3$) produced during the evaporation of the PBH. Once the number of DM particles produced from PBH evaporation is known, the DM relic abundance $\Omega_\text{DM}h^2$ at the present epoch can be calculated as

\begin{equation}
    \Omega_\text{DM}\,h^2 =\mathbb{C}(T_\text{ev})
    \begin{cases}
      \frac{1}{\pi^2}\,\sqrt{\frac{M_\text{pl}}{m_\text{in}}}\,m_\text{DM} &\text{for } m_\text{DM} < T_\text{BH}^\text{in}\,,\\
       \frac{1}{64\,\pi^4}\left(\frac{M_\text{pl}}{m_\text{in}}\right)^{5/2}\,\frac{M_\text{pl}^2}{m_\text{DM}} &\text{for} ~m_\text{DM} > T_\text{BH}^\text{in},
    \end{cases}\label{eq:rel-dm}\,
\end{equation}

\noindent with $\mathbb{C}(T_\text{ev})=\frac{s_0}{\rho_c}\,\frac{1}{\zeta}\,\frac{g_{X,H}}{g_{\star,H}}\,\frac{5}{g_{\star s}(T_\text{ev})}\,\,\left(\frac{\pi^3\,g_{\star}(T_\text{ev})}{5}\right)^{3/4}\,\sqrt{\frac{\mathcal{G}\,g_{\star,H}}{10640\,\pi}}$. Here $\zeta$ parametrizes a possible entropy production after PBH evaporation until now, i.e., $\zeta\,\left(sa^3\right)_\text{ev}=\left(sa^3\right)_0$. Looking at the above equation one finds that the heavier the DM mass (for $m_\text{DM} > T_\text{BH}^\text{in}$), the lesser will be the DM relic abundance. On contrary, in a scenario with $m_\text{DM} < T_\text{BH}^\text{in}$, the DM relic abundance becomes proportional to the DM mass. The parameter space for DM produced during PBH evaporation is highly constrained. The correct relic abundance is satisfied only if $m_\text{DM}\gtrsim 10^{10}$ GeV and $m_\text{in}\gtrsim 10^6$g~\cite{Fujita:2014hha, Samanta:2021mdm}. In the rest of the parameter space, the DM is mostly overabundant if one assumes that there exists no entropy injection into the thermal plasma after the evaporation of the PBH, $i.e.~\zeta = 1$. This tension can be relaxed considerably if one assumes $\zeta\sim\mathcal{O}(10)~\text{or}~\mathcal{O}(100)$. While DM of mass around keV or lighter do not get overproduced, it suffers from strong Lyman-$\alpha$ bounds \cite{Irsic:2017ixq, Diamanti:2017xfo, Bernal:2020bjf} severely restricting the parameter space in $\beta-m_{\rm in}$ plane. We do not consider such light or superheavy DM in our setup and focus on the intermediate mass range which typically gets overproduced during a PBH dominated phase.

%%%%%%%%%%%%%%%%%%%%%%%%%%%
\begin{figure}[htb!]
$$
\includegraphics[scale=0.15]{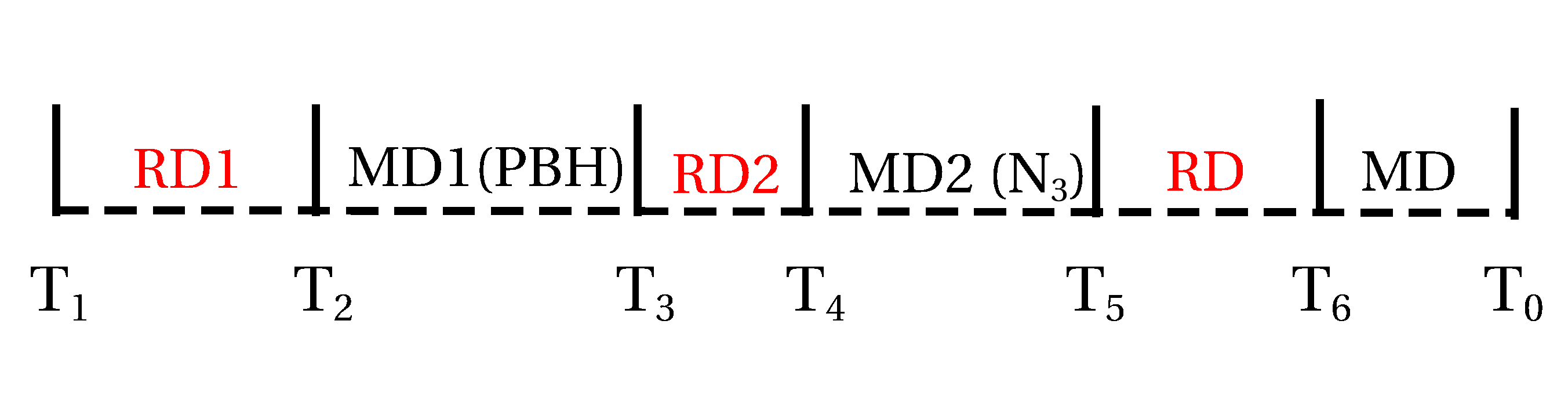}
$$
\caption{A schematic of the dominant phases of evolution of the universe at different epochs from a reheating temperature $T_1$ to the present temperature $T_0$ (neglecting recent dark energy domination). $T_3$ is the black hole evaporation temperature (equation \eqref{eq:pbh-Tev}) and $T_6$ represents the standard matter-radiation equality temperature of $\sim 0.75$ eV.}
\label{fig:schematic}
\end{figure}

%%%%%%%%%%%%%%%%%%%%%%%%%%%%

%%%%%%%%%%%%%%%%%%%%%%%%%%%
\begin{figure}[htb!]
$$
\includegraphics[scale=0.45]{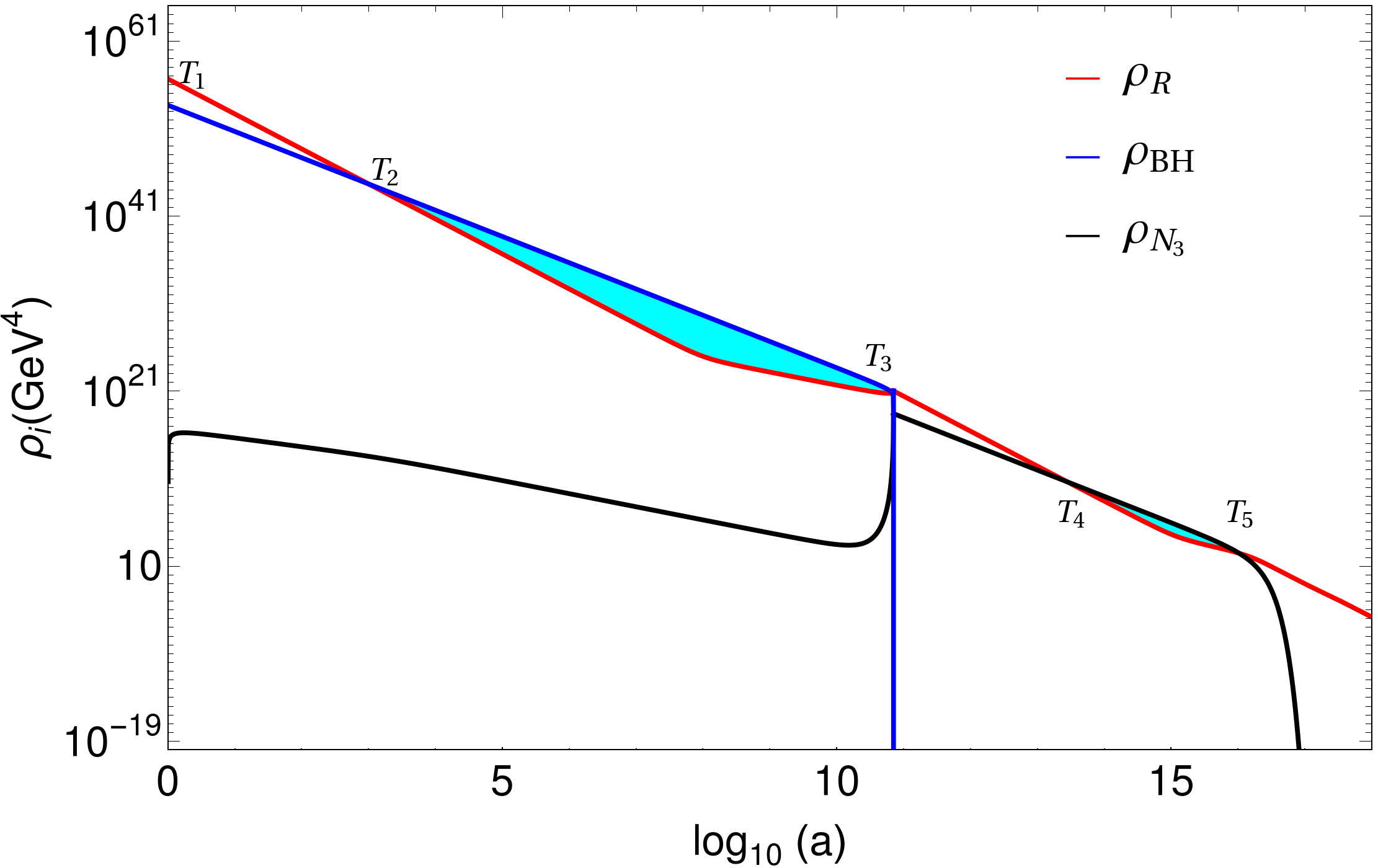}
$$
\caption{Evolution of the energy densities of different components in the universe clearly showing the two intermediate early matter-dominated phases due to PBH and diluter $N_3$ respectively. $a$ represents the scale factor with the initial value $a_{\rm in}=1$. Here, we take $m_{\rm in}=5000$ g, $M_{\rm DM}=0.1$ GeV, $M_{3}=10^{11}$ GeV.}
\label{fig:rho}
\end{figure}

%%%%%%%%%%%%%%%%%%%%%%%%%%%%

Analogous to the DM, the diluter $N_3$ is also produced during the PBH's evaporation whose abundance can be calculated using Eq.~\eqref{particle_yield}. Once produced, it decays very slowly as a result of its very feeble interaction with the first generation of the SM lepton doublet and the Higgs. Subsequent to the PBH evaporation, the  universe enters into a radiation-dominated era for a short duration following which the energy density of $N_3$ overtakes the radiation energy density and the universe re-enters the matter-dominated phase in a similar manner as was shown in~\cite{Barman:2022gjo}. The universe experiences a change in temperature evolution and cools slower as the diluter starts to decay. This is because, in the process of its decay, the diluter also injects entropy into the thermal bath. Finally, after the decay of the $N_3$ is complete the universe again becomes radiation-dominated. A schematic of the scenario is shown in Fig. \ref{fig:schematic} and the evolution of the energy densities of the different components is shown in Fig. \ref{fig:rho} for a particular choice of benchmark parameters.  As a result of the entropy injection, the number density of DM is brought within observed limits thereby alleviating the tension of DM relic abundance produced from PBH evaporation in the intermediate mass range mentioned before. 

It should be noted that in addition to PBH evaporation, DM and diluter can also be produced from $2$ to $2$ scatterings mediated by massless graviton. Adopting the approach discussed in earlier works \cite{Barman:2022gjo,Garny:2015sjg,Tang:2017hvq,Garny:2017kha, Bernal:2018qlk,Barman:2021ugy,Bernal:2021yyb}, we find that such productions are negligible for the choices of $m_{\rm in}, M_{\rm DM}$ and $N_3$ mass $M_3$ considered in our numerical analysis to be discussed below.

Finally, we would like to mention here that one can also have leptogenesis from the decay of the RHNs produced from PBH evaporation. However, as studied in details in Refs. \cite{Datta:2020bht, JyotiDas:2021shi, Barman:2021ost, Barman:2022gjo}, the baryon asymmetry produced for the PBH mass range considered in our analysis, the asymmetry turns out to be smaller than the observed value. As we will see in the next section, decreasing the PBH mass is not favorable from the gravitational wave detection perspective as it would increase the turning point frequency to a higher value beyond the reach of near-future detectors and would also result in particle production from cosmic strings to be dominant compared to gravitational wave emission. Moreover, the asymmetry that might be produced thermally at a high temperature would be diluted because of two epochs of entropy dilution arising from PBH evaporation and $N_3$ decay.

%%%%%%%%%%%%%%%%%%%%%%%%%%%%%%%%%%%%%%%%%%%%%%%%%%%%%%%%%%%%%%%%%%%%%%%%%%%%%%%%%%%%%%%%%%%%%%%%%%%%%%%%%%%%%%%%%%%%%%%%%%%%%%%%%%%%%%%%%
\section{Gravitational Waves from Cosmic Strings}
\label{sec3}
Cosmic strings \cite{Kibble:1976sj, Nielsen:1973cs}, are one-dimensional objects which appear as topological defects after the spontaneous breaking of a symmetry group containing a vacuum manifold that is not simply connected \cite{Nielsen:1973cs}. The simplest group that exhibits such a feature is $U(1)$ which naturally appears in many BSM frameworks. Numerical simulations \cite{Ringeval:2005kr,Blanco-Pillado:2011egf} based on Nambu-Goto action indicate that dominant energy loss from a string loop is in the form of GW radiation if the underlying symmetry is gauged. Thus, being one of the potential sources of primordial GW, they have gained a great deal of attention after the recent finding of a stochastic common spectrum process across many pulsars\cite{NANOGrav:2020bcs,Goncharov:2021oub,Ellis:2020ena,Blasi:2020mfx,Samanta:2020cdk}.  If the symmetry breaking scale ($\Lambda_{CS}$) is high enough ($\Lambda_{CS}\gtrsim 10^9$ GeV), the resulting GW background is detectable. This makes CS an outstanding probe of super-high scale physics\cite{Buchmuller:2013lra, Dror:2019syi, Buchmuller:2019gfy,King:2020hyd, Fornal:2020esl,  Buchmuller:2021mbb, Masoud:2021prr,Lazarides:2021uxv, Afzal:2022vjx, Borah:2022byb,Lazarides:2022jgr,Lazarides:2022spe,Maji:2022jzu}. The properties of CS are described by their normalised tension $G \mu\sim G \Lambda_{\rm CS}^2$ with $G$ being the Newton's constant. Unless the motion of a long-string network gets damped by thermal friction\cite{Vilenkin:1991zk}, shortly after formation, the network oscillates ($t_{\rm osc}$) and  enters the scaling regime\cite{Blanco-Pillado:2011egf,Bennett:1987vf,Bennett:1989ak} which is an attractor solution of two competing dynamics--stretching of the long-string correlation length due to cosmic expansion and fragmentation of the long strings into loops which oscillate to produce particle radiation or GW\cite{Vilenkin:1981bx,Turok:1984cn,Vachaspati:1984gt}. 

A set of normal-mode oscillations with frequencies $f_k=2k/l$ constitute the total energy loss from a loop, where the mode numbers $k=1,2,3....\infty$ \footnote{An upper bound on
k appears from the violation of the Nambu-Goto approximation for high k values \cite{Gouttenoire:2019kij}.}. Therefore, the GW energy density parameter is defined as $\Omega_{\rm GW}(t_0,f)=\sum_k\Omega_{\rm GW}^{(k)}(t_0,f)$, with $t_0$ being the present time and $f\equiv f(t_0)= f_k a(t_0)/a(t)$. Present day GW energy density corresponding to the mode $k$ is computed with the integral \cite{Blanco-Pillado:2013qja} 
\begin{align}
 \Omega_{\rm GW}^{(k)}(t_0,f)=\frac{2k G\mu^2\Gamma_k}{f \rho_c}\int_{t_{osc}}^{t_0}dt\left[\frac{a(t)}{a(t_0)}\right]^5 n\left(t,l_k\right),\label{gwint}
\end{align}
where $n\left(t,l_k\right)$ is a scaling loop number density which can be computed analytically using Velocity-dependent-One-Scale (VOS) \cite{Martins:1996jp,Martins:2000cs,Auclair:2019wcv} model \footnote{Compared to the numerical simulation, VOS model overestimates loop number density by a factor of 10. We, therefore, use a normalization factor $\mathcal{F}_\alpha=0.1$ to be consistent with simulation \cite{Auclair:2019wcv}.}, $\rho_c$ is the critical energy density of the universe and $\Gamma_k=\frac{\Gamma k^{-\delta}}{\zeta(\delta)}$ depends on the small scale structures in the loops such as cusps ($\delta=4/3$) and kinks ($\delta=5/3$). In this article, we consider only cusps to compute GW spectrum. A similar analysis can also be done considering kinks. Although for higher modes, contributions to the GWs from cusps are dominant, the number of cusps and kinks per loop oscillation cannot be straightforwardly determined with numerical simulations, which do not include gravitational wave back-reaction in general. A preference for cusps over kinks comes from the so-called smoothing mechanism, where the kinky loops are expected to be smoothed out due to the gravitational wave back-reaction \cite{Blanco-Pillado:2015ana}. This mechanism, however, has been challenged in \cite{Wachter:2016hgi,Wachter:2016rwc}. Therefore, we would like to place the consideration of cusps over kinks in this article as a choice rather than a preference. 

A typical feature of GWs from CS is a flat plateau due to loop formation and decay during radiation domination, with an amplitude given by 
\begin{align}
\Omega_{\rm GW}^{(k=1),{\rm plateau}}(f)=\frac{128\pi G\mu}{9\zeta(\delta)}\frac{A_r}{\epsilon_r}\Omega_r\left[(1+\epsilon_r)^{3/2}-1\right], \label{flp1}
\end{align}
where $\epsilon_r=\alpha/\Gamma G\mu$ with $\alpha$ the initial (at $t=t_i$) loop size parameter, $\Omega_r\simeq 9\times 10^{-5}$ and $A_r=5.4$ \cite{Auclair:2019wcv}. In our analysis, we have considered $\alpha \simeq 0.1$\,\cite{Blanco-Pillado:2013qja,Blanco-Pillado:2017oxo} (as suggested by simulations),  $\Gamma\simeq 50$\,\cite{Vachaspati:1984gt,Blanco-Pillado:2013qja,Blanco-Pillado:2017oxo}, and CMB constraint $G\mu\lesssim 10^{-7}$ \cite{Charnock:2016nzm} which lead to $\alpha \gg \Gamma G \mu$. In this limit, Eq.\eqref{flp1} implies $\Omega_{\rm GW}^{(k=1)}(f)\sim \Lambda_{\rm CS}$, a property that makes models with larger breaking scales more testable with GWs from CS. The observed frequency today of the GW spectrum is connected to the temperature at which the loops responsible for that particular frequency have been formed. The major contribution to the GW emission is at a time when the loop reaches its half-life $t_E = t_H \simeq \frac{\alpha t_i}{2 \Gamma G \mu }$, where $t_i$ indicates the time of loop formation. Now, since the frequency emitted at the time $t_E$ is related to the length of the loop as $\left(f (t_E)= 2k / l (t_E)\right)$, we get (considering k=1)
\begin{align}
    \alpha t_i & \simeq \frac{4}{f}\frac{a(t_E)}{a(t_0)} \nonumber \\
                & \simeq \frac{4}{f}\left(\frac{t_E}{t_{\rm eq}}\right)^{1/2}\left(\frac{t_{\rm eq}}{t_{0}}\right)^{2/3}\nonumber \\
      \implies  f &\simeq \sqrt{\frac{8 z_{\rm eq}}{\alpha\Gamma G\mu}}\left(\frac{t_{\rm eq}}{t_{i}}\right)^{1/2} t^{-1}_0 \,,\label{eqn:fhlflf}
\end{align}
where $t_{\rm eq}$ ($z_{\rm eq}$ ),  $t_{0}$ indicate the standard matter-radiation equality time (redshift) and the present time respectively.

The dependence of the GW amplitude on the frequency can be qualitatively understood as follows \cite{Gouttenoire:2019kij}. The frequency dependence receives contribution from two factors. Because of the Universe's expansion, the GW  energy density dilutes as $a^{-4}$.  Hence, the GW amplitude emitted by loops formed at higher temperatures and subsequently at higher frequencies undergoes more suppression until the present time.  However, at earlier times the loop production rate, which varies as $\propto t_i^{-4}$ \cite{Blanco-Pillado:2013qja, Cui:2017ufi} also increases, leading to enhancement of the GW amplitude.  Thus, the total dependence on frequency is the effect of these two factors. In the case of radiation domination, these two contributions exactly cancel each other, leading to a flat spectrum. If there exist cosmological epochs other than radiation at higher temperatures, there would be a deviation from flatness, which would first start to appear because of the contribution of those loops that have been formed at a time say $t_{\Delta}$ (with corresponding temperature $T_\Delta$), when the standard radiation era began or the non-standard era ended. Thus, the frequency at which the spectral break occurs or the turning point frequency $f_\Delta$ is given by Eq. \eqref{eqn:fhlflf} with $t_i = t_\Delta$ \cite{Cui:2017ufi,Cui:2018rwi,Gouttenoire:2019kij}, which in terms of temperature gives us 
\begin{align}
 f_\Delta \simeq \sqrt{\frac{8}{z_{\rm eq}\alpha\Gamma G\mu}}\left(\frac{g_*(T_\Delta)}{g_*(T_0)}\right)^{1/4}\frac{T_\Delta}{T_0}t_0^{-1}\,.\label{eq:TPF}
\end{align}
Beyond $f_\Delta$, the spectrum goes as $\Omega_{\rm GW}\sim f^{-1}$ for $k=1$ mode, which can be estimated by considering the corresponding change in the loop number density and the cosmic evolution because of the matter-dominated background (Please refer to \cite{Gouttenoire:2019kij} for a detailed derivation). 

% When infinite modes are summed, $\Omega_{\rm GW}\sim f^{-1/3}$\cite{Blasi:2020wpy,Gouttenoire:2019kij,Datta:2020bht, Gouttenoire:2019rtn}}\footnote{\textcolor{blue}{The spectrum for the $k^{\rm th}$ mode is related to the fundamental $k=1$ mode through (cf. Eq. \eqref{gwint}) \cite{Gouttenoire:2019kij} 
% \begin{align}
% \Omega_{GW}^{(k)} (f)=k^{-4/3}\Omega_{GW}^{(1)} (f/k).
% \end{align}
% For the part of GW spectra varying as $\Omega_{GW}^{(1)} \propto f^{-1}$, summing over all modes changes this slope to $f^{-1/3}$ for early matter domination \cite{Gouttenoire:2019kij}.}}. 

Now, as shown in Fig. \ref{fig:schematic}, in our scenario we have two phases of early matter domination, one because of $N_3$ domination just before BBN (MD2) and the other because of PBH (MD1) at much earlier epochs. This leads to some interesting patterns in the GW spectrum as we will show below. First, let us discuss the impact of the recent matter domination because of $N_3$ (MD2), which ends at temperature $T_5$. Note that the lifetime of $N_3$ should be large enough such that it does not decay before leading to matter-domination. We attempt to find an analytical estimate for $T_5$ in terms of the other parameters in our scenario like $M_{\rm DM}$, $m_{\rm in}$, $M_3$ such that the turning point frequency $f_\Delta$ given by equation \eqref{eq:TPF} (with $T_\Delta = T_5$), can be related to these parameters. For this purpose, we need to calculate the entropy dilution arising because of $N_3$ decay, given by $S= \frac{s(T_5) a(T_5)^3}{s(T_4) a(T_4)^3}$, which can be estimated to be \cite{Kolb:1990vq, Scherrer:1984fd, Bezrukov:2009th} 
\begin{align}
 S\simeq \left[2.95\times \left(\frac{2\pi^2 g_{*}(T_{N_3})}{45}\right)^{1/3}\frac{(Y_{N_3} M_{3})^{4/3}}{(\Gamma_{3} M_{\rm pl})^{2/3}}\right]^{3/4},\label{eq:dilutionF}
\end{align}
where $Y_3=n_{N_3}/s$ is the initial yield of $N_3$ arising from PBH and $\Gamma_3$ is its decay width. This should be equal to $=\Omega_{\rm DM}h^2/0.12$ to produce the observed DM abundance. $T_{N_3}\simeq T_5$ represents the temperature at the end of $N_3$ decay and assuming instantaneous decay of $N_3$, we write 
\begin{align}
    \Gamma_{3} M_{\rm pl} = 1.66 \sqrt{g_{*}(T_{N_3})} T_{N_3}^2. \label{eq: N3dec}
\end{align}
Note that $Y_{N_3}$, in turn depends on $m_{in}$, $M_3$ (cf. Eq. \eqref{particle_yield}). Using Eq.  \eqref{particle_yield} for $Y_{N_3}$, Eq. \eqref{eq:rel-dm} for $\Omega_{DM}h^2$, Eq. \eqref{eq: N3dec} for $\Gamma_{3}$  and substituting in Eq. \eqref{eq:dilutionF}, we find 
\begin{equation}
   T_{N_{3}}\simeq T_5 \simeq 
    \begin{cases}
     4\times 10^{16} \left(\frac{1~ \text{GeV}}{M_{\rm DM}}\right) \left(\frac{1 ~\text{GeV}}{M_{3}}\right)\left(\frac{1 ~g}{m_{\rm in}}\right)^2  \text{GeV} &\text{for } M_3 > T_\text{BH}^\text{in}\,,\\[8pt]
        3.6 \times 10^{-10}\frac{M_3}{M_{\rm DM}} \text{GeV} &\text{for } M_3 < T_\text{BH}^\text{in}\,.
    \end{cases}\label{eq:T5}\,,
\end{equation}

For both the cases above, $T_5$ decreases with a higher value of dark matter mass $M_{\rm DM}$. This is expected since the dark matter relic from PBH ($M_{\rm DM}<T_{\rm BH}^{\rm in}$) increases with $M_{\rm DM}$ and hence a longer duration of $N_3$ domination would be required to produce the observed relic, which decreases $T_5$. Now, for the case $M_3> T_{\rm BH}^{\rm in}$, increasing the initial PBH mass decreases the DM relic (see Eq. \eqref{eq:rel-dm}), and also the $N_3$ yield, with the latter effect being more dominant. Hence, $N_3$ starts dominating much later which results in a smaller $T_5$. For the other case $M_3< T_{\rm{BH}}^{\rm in}$, the dependence on $m_{\rm in}$ does not appear because of a similar dependence of the $N_3$ yield on $m_{\rm in}$ as that of DM. Finally, increasing the diluter mass $M_3$ decreases its yield for $M_3>T_{\rm{BH}}^{\rm in}$ and leads to a decrease in $T_5$. The reverse is true for the case $M_3<T_{\rm{BH}}^{\rm in}$. In our analysis, we consider $M_3 > T_{\rm{BH}}^{\rm in}$, and fix its value at $M_3=10^{11}~ \rm GeV$\footnote{The initial PBH mass $m_{\rm in}$ should be high enough (and hence a lower $T_{\rm{BH}}^{\rm in}$) for the tilt related to PBH evaporation to be within sensitivity.}. Thus, the turning point frequency (Eq. \eqref{eq:TPF}) varies as $f_\Delta \propto \frac{1}{M_{\rm DM}}\frac{1}{m_{\rm in}^2}$.

%%%%%%%%%%%%%%%%%%%%%%%%%%%%%%%%%%%%%%%%%%%
\begin{figure}[htb!]
$$
\includegraphics[scale=0.45]{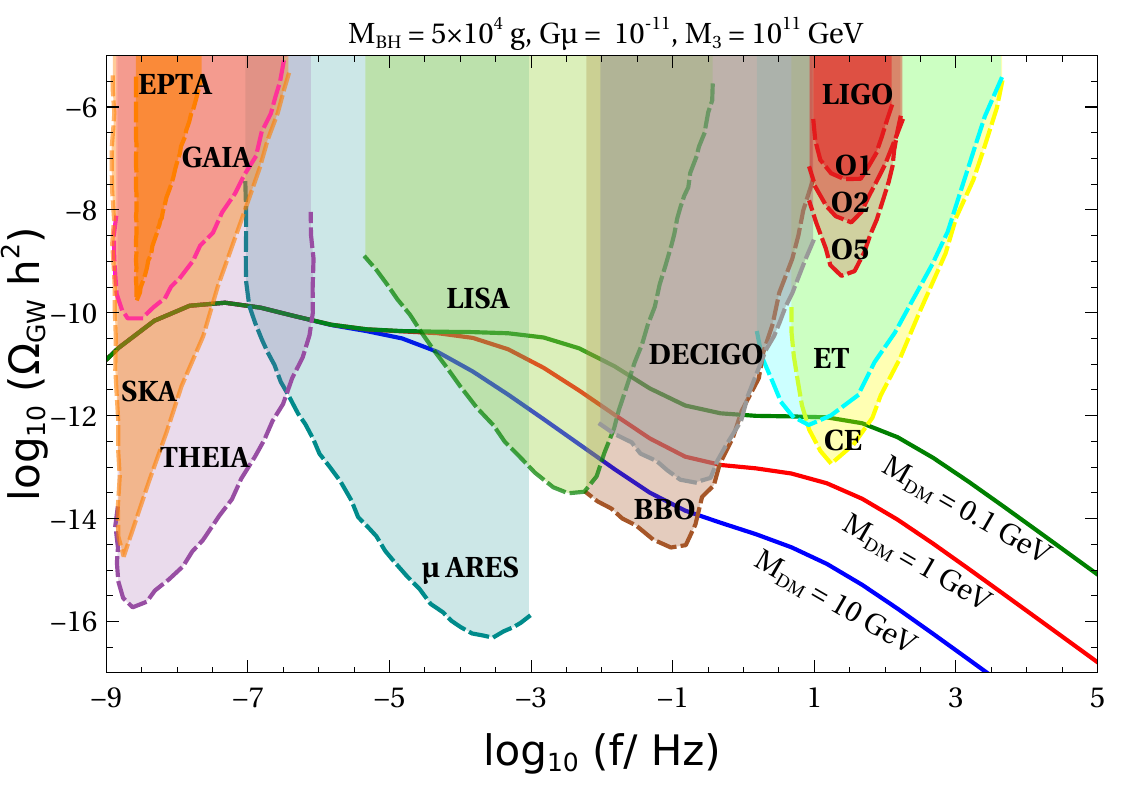}~~~~\includegraphics[scale=0.45]{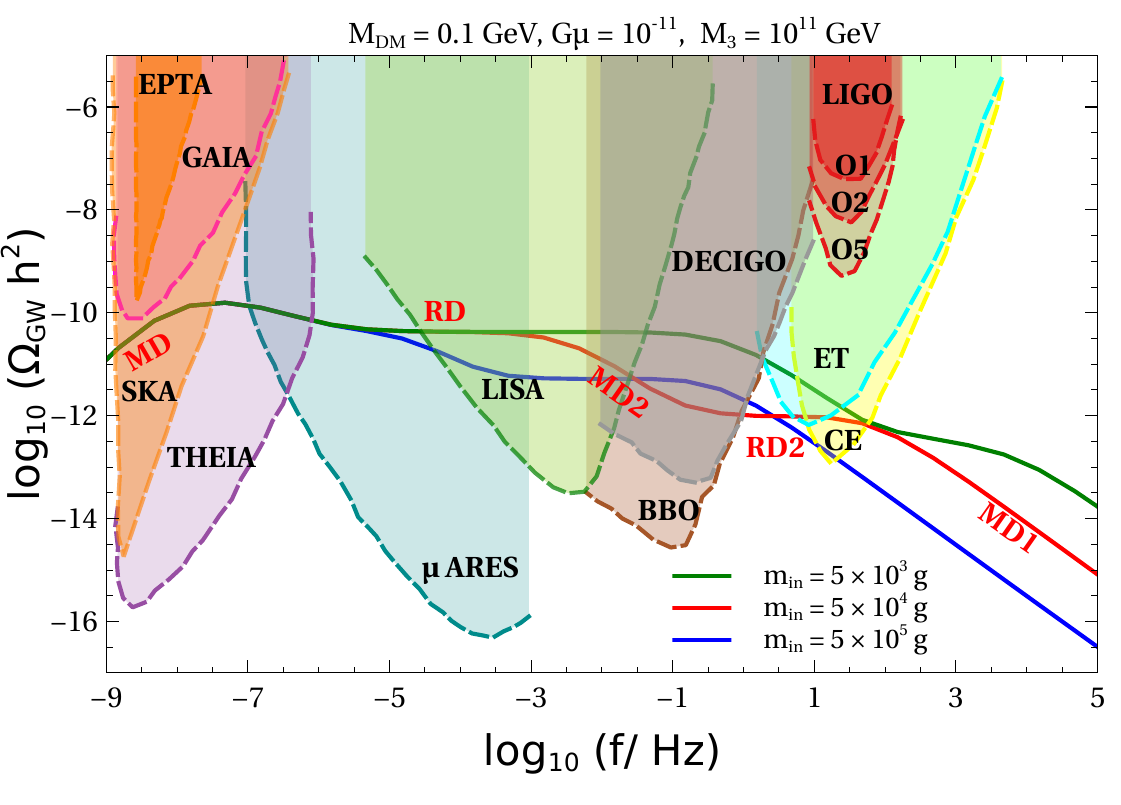}
$$
\caption{Gravitation wave spectra for the fundamental ($k=1$) mode for different values of DM mass (left panel) and initial PBH mass (right panel), satisfying the condition $\Omega_{\rm DM}h^2=0.12$. The experimental sensitivities of SKA \cite{Weltman:2018zrl}, GAIA \cite{Garcia-Bellido:2021zgu}, EPTA \cite{Kramer_2013}, THEIA \cite{Garcia-Bellido:2021zgu}, $\mu$ARES \cite{Sesana:2019vho}, LISA\,\cite{2017arXiv170200786A}, DECIGO \cite{Kawamura:2006up}, BBO\,\cite{Yagi:2011wg}, ET\,\cite{Punturo_2010}, CE\,\cite{LIGOScientific:2016wof} and aLIGO \cite{LIGOScientific:2014pky} are shown as shaded regions of different colours. The different cosmic regimes of Fig. \ref{fig:schematic} are indicated for one of the BPs in the right panel (see text).}
\label{fig:GW_spec1}
\end{figure}
%%%%%%%%%%%%%%%%%%%%%%%%%%%%%%%%%%%%%%%%%%%%

Next, we look into the effects of the matter-dominated era arising from PBH (MD1). Here, the expression for the turning point frequency $f_{\Delta}$ given by Eq. \eqref{eq:TPF} is changed because of the other intermediate epochs before BBN. This can be understood as follows. A loop formed at the time $t_3$ would contribute to the spectrum much later when it reaches its half-life which is given by \cite{Gouttenoire:2019kij} 
\begin{align}
    \tilde{t}_M \simeq \frac{\alpha t_3}{2 \Gamma G \mu } \label{eq:halflf} 
\end{align}
The frequency emitted is related to the length of the loop at its half-life time $\left(f (\tilde{t}_M)= 2k / l (\tilde{t}_M)\right)$ and hence we get
\begin{align}
 \alpha t_3 \simeq \frac{4}{f}\frac{a(\tilde{t}_M)}{a(t_0)}  
\end{align}
where $f$ is the frequency observed today. Now, usually, $\tilde{t}_M$ is in the radiation-dominated era after BBN, which gives us Eq. \eqref{eq:TPF}. However, in our case, $\tilde{t}_M$ would be in the radiation or early matter-dominated era before BBN, which gives us\footnote{For most of our parameters used, the radiation-dominated era after PBH domination (RD2) lasts for a very small duration. Hence, $\tilde{t}_M$ is in the $N_3$ dominated era. However, for large PBH masses, the radiation era is prolonged and loops reach their half-life $\tilde{t}_M$ in the radiation era itself, which modifies Eq. \eqref{eqn:lngth_hl} as $\alpha t_3 \simeq \frac{4}{f} \left(\frac{\tilde{t}_M}{t_4}\right)^{1/2} \left(\frac{t_4}{t_5}\right)^{2/3} \left(\frac{t_5}{t_6}\right)^{1/2} \left(\frac{t_6}{t_0}\right)^{2/3}$.}

\begin{align}
   \alpha t_3 \simeq  \frac{4}{f} \frac{a(\tilde{t}_M)}{a(t_5)} \frac{a(t_5)}{a(t_6)} \frac{a(t_6)}{a(t_0)}\simeq \frac{4}{f} \left(\frac{\tilde{t}_M}{t_5}\right)^{2/3} \left(\frac{t_5}{t_6}\right)^{1/2} \left(\frac{t_6}{t_0}\right)^{2/3}\label{eqn:lngth_hl}
\end{align}
Substituting $\tilde{t}_M$ from Eq. \eqref{eq:halflf}, we find the turning point frequency, which we call $f'_{\Delta}$ to be given by
\begin{align}
f'_{\Delta}\simeq \frac{4}{\alpha^{1/3} t_3^{1/3}t_5^{2/3}\left( 2 \Gamma G \mu \right)^{2/3}}\frac{T_{6}}{T_5}\left(\frac{g_{*s}(T_6)}{g_{*s}(T_5)}\right)^{1/3} z_{\rm eq}^{-1}
\end{align}
Using $t_{3,5}\sim \frac{1}{2 T_{3,5}^2}$, we get
\begin{align}
f'_{\Delta} \propto T_{3}^{2/3} T_{5}^{1/3}. \label{eqn:TPF2}
\end{align}
Thus, the TPF $f'_{\Delta}$ depends not only on the PBH evaporation temperature $T_3$, but also on $T_5$, i.e, the temperature when $N_3$ dominated era ends. Using Eq. \eqref{eq:T5} and \ref{eq:pbh-Tev}, we find  $f'_\Delta \propto \frac{1}{M_{\rm DM}^{1/3}} \frac{1}{m_{\rm in}^{5/3}}$.

Similar turning point frequencies can also be obtained for loops below a critical length $l_{c}=\frac{\mu^{-1/2}}{\left (\Gamma G\mu \right )^2}$ \cite{Auclair:2019jip}, for which particle production becomes dominant over the GW emission. This translates into the bound $G\mu>2.4 \times 10^{-16} T_{\Delta}^{4/5}$. In our scenario, we choose $G\mu=10^{-11}$ which is high enough such that the GW emission always dominates. This gives an upper bound on the TPF as $T_{\Delta}\lesssim 5\times10^{5}$ GeV, and considering $T_{\Delta}=T_3=T_{\rm ev}$, we get a lower bound on the initial PBH mass as $m_{\rm in}\gtrsim2\times 10^{3}$ g.

%%%%%%%%%%%%%%%%%%%%%%%%%%%%%%%%%%%%%%%%%%%
%\begin{figure}[htb!]
%$$
%\includegraphics[scale=0.5]{high_k.pdf}
%$$
%\caption{Effect of summing over the high k modes on the GW spectrum.}
%\label{fig:highk}
%\end{figure}
%%%%%%%%%%%%%%%%%%%%%%%%%%%%%%%%%%%%%%%%%%%%

%%%%%%%%%%%%%%%%%%%%%%%%%%%%%%%%%%%%%%%%%%%
\begin{figure}[htb!]
$$
\includegraphics[scale=0.5]{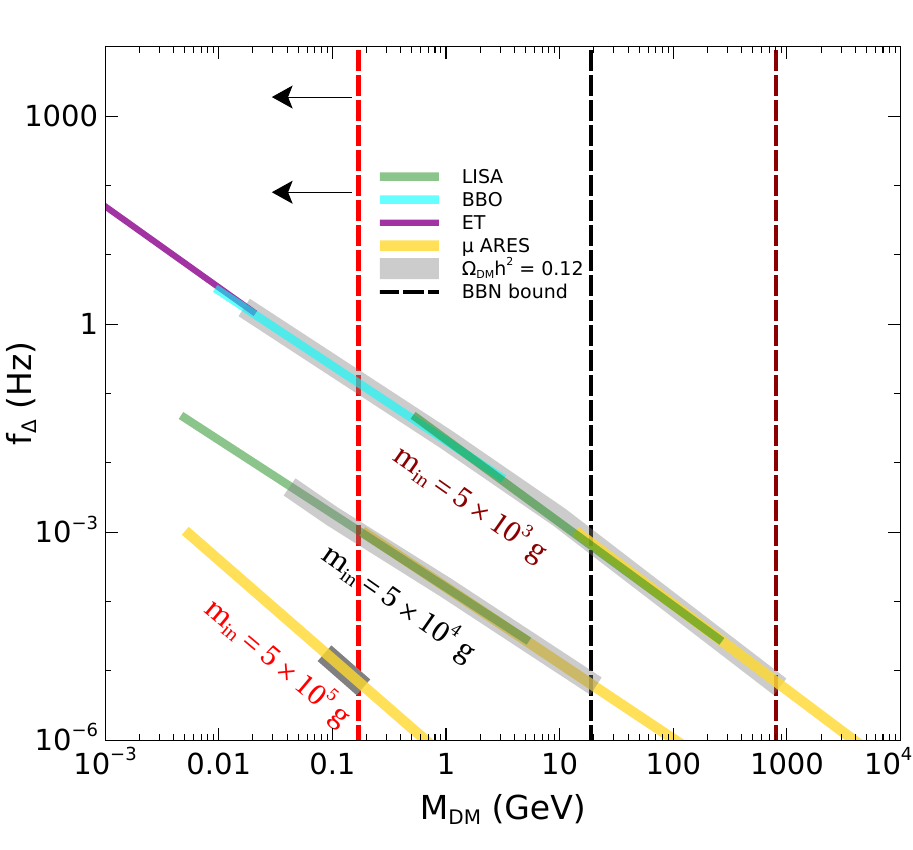}~~~~\includegraphics[scale=0.5]{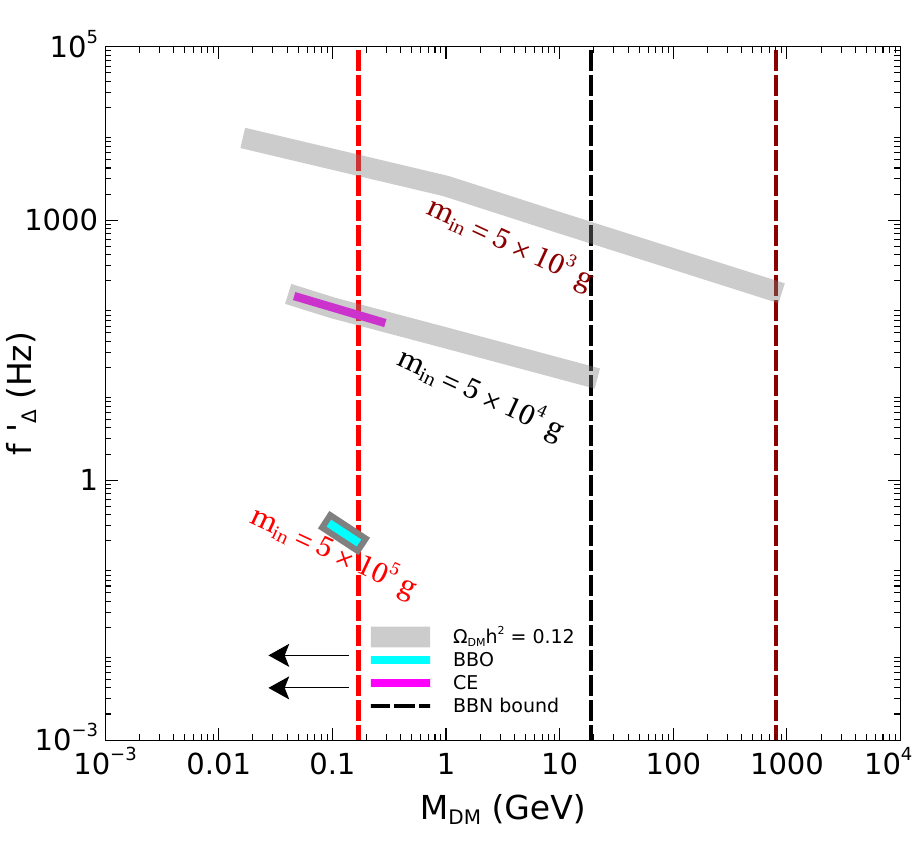}
$$
\caption{Contours satisfying $\Omega_{\rm DM}h^2=0.12$ in the $M_{\rm DM}-f_{\Delta}$ plane (left panel) and the $M_{\rm DM}-f'_{\Delta}$ (right panel), along with the sensitivities of future experiments like LISA\,\cite{2017arXiv170200786A}, BBO\,\cite{Yagi:2011wg}, ET\,\cite{Punturo_2010}, CE\,\cite{LIGOScientific:2016wof}, $\mu$ARES \cite{Sesana:2019vho}. Here, $G\mu=10^{-11}$ and $M_3=10^{11}$ GeV.}
\label{fig:MDM_TPF}
\end{figure}
%%%%%%%%%%%%%%%%%%%%%%%%%%%%%%%%%%%%%%%%%%%%

We now compute the GW spectra by integrating Eq. \eqref{gwint} over the different cosmological epochs. All the temperatures ($T_2$, $T_3$, $T_4$, $T_5$) corresponding to the different cosmological backgrounds (see Fig. \ref{fig:schematic}) which would be required in this integration are found by numerically solving the following coupled Boltzmann equations \cite{Perez-Gonzalez:2020vnz, JyotiDas:2021shi, Barman:2021ost} for $\rho_{\rm BH}$, $\rho_{\rm R}$, $\rho_{\rm DM}$ and $\rho_{\rm N_{3}}$
\begin{align}
% \begin{split}
&  \frac{dm_\text{BH}}{da} = -\frac{\kappa}{a\,\mathcal{H}}\,\epsilon(m_\text{BH})\,\left(\frac{1\text{g}}{m_{\rm BH}}\right)^2\,,
\nonumber\\&
\frac{d\widetilde{\rho}_R}{da}=-\frac{\epsilon_\text{SM}(m_\text{BH})}{\epsilon(m_\text{BH})}\,\frac{a}{m_\text{BH}}\,\frac{dm_\text{BH}}{da}\,\widetilde{\rho}_\text{BH} + \frac{a}{\mathcal{H}} \Gamma_{3} M_{3}\widetilde{n}_{N_3}^\text{BH}\,,
\nonumber\\& 
\frac{d\widetilde{\rho}_\text{BH}}{da}=\frac{1}{m_\text{BH}}\,\frac{dm_\text{BH}}{da}\,\widetilde{\rho}_\text{BH}\,,
\nonumber\\&
a\mathcal{H}\frac{d\widetilde{n}_{N_3}^\text{BH}}{da}=\Gamma_{\text{BH}\to N_3}\,\frac{\widetilde{\rho}_{BH}}{m_{BH}} -\Gamma_{3}\widetilde{n}_{N_3}^\text{BH}
\nonumber\\&
a\mathcal{H}\frac{d\widetilde{n}_{DM}^\text{BH}}{da}=\Gamma_{\text{BH}\to DM}\,\frac{\widetilde{\rho}_{BH}}{m_{BH}}
\nonumber\\&
\frac{dT}{da}=-\frac{T}{\Delta}\Biggl[\frac{1}{a}+\frac{\epsilon_\text{SM}(m_\text{BH})}{\epsilon(m_\text{BH})}\,\frac{1}{m_\text{BH}}\,\frac{dm_\text{BH}}{da}\,\frac{g_\star(T)}{g_{\star s}(T)}\,a\,\frac{\widetilde{\rho}_{BH}}{4\,\widetilde{\rho}_{R}}+\frac{\Gamma_{3}M_{3}}{3 \mathcal{H} ~s~ a^4}\widetilde{n}_{N_3}^\text{BH}\Biggr]\,.    
\end{align}
Here, $\Delta = 1 + \frac{T}{3 g_{* s}(T)} \frac{d g_{* s}(T)}{d T}$. $\widetilde{\rho}_{R}$ and $\widetilde{\rho}_\text{BH}$ denote the comoving energy densities of radiation and PBH respectively, $\mathcal{H}$ is the Hubble parameter, whereas $\widetilde{n}_{N3}$ and $\widetilde{n}_{DM}$ represent the comoving number densities of $N_3$ and DM respectively. Note that the terms on the R.H.S of $\widetilde{\rho}_{R}$ evolution equation denote entropy dilution from PBH evaporation and $N_3$ decay, with the corresponding second and third terms in the temperature evolution equation. Since we are considering a scenario of DM with only gravitational interactions, it gets produced only from PBH. Similarly, production of $N_3$ from the thermal bath also remains suppressed because of its tiny Yukawa couplings required for a longer lifetime. $\Gamma_{\text{BH}\to N_3}$, $\Gamma_{\text{BH}\to \text{DM}}$ denote the production rate from PBH \cite{Perez-Gonzalez:2020vnz}, whereas $\Gamma_{3}$ indicates the decay rate of $N_3$.

In Fig. \ref{fig:GW_spec1}, we show the GW spectra for three DM masses (left panel) and three PBH masses (right panel), differing by orders of magnitude. The GW spectral behavior can be connected to the different cosmic regimes represented in Fig. \ref{fig:schematic}. We show this for one of the BPs in the right panel of Fig.  \ref{fig:GW_spec1}. Here, the label indicates the contribution to the GW spectrum from loops emitting in different cosmic eras. We refrain from showing the contribution coming from RD1 since it occurs at ultra high frequencies with sub-dominant amplitude out of reach of near-future detectors. The decay rate of $N_3$ is chosen such that $\Omega_{\rm DM}h^2=0.12$. As anticipated above, the spectral breaks due to the two early matter-dominated eras can be clearly seen. The first break ($f_\Delta$) at the lower frequency, from flat to power-law, depends on the temperature $T_5$ which marks the end of $N_3$ domination. It is evident here that $f_\Delta$ decreases with DM mass $M_{\rm DM}$ and initial PBH mass $m_{\rm in}$. The break from flat to power-law at higher frequencies ($f'_{\Delta}$) is related to PBH evaporation at the temperature $T_3$, and a higher evaporation temperature (lower $m_{\rm in}$) implies break at a higher frequency (right panel). However, as discussed above, this break also depends on the other intermediate eras before BBN, specifically on the temperature $T_{5}$ (see Eq. \eqref{eqn:TPF2}). This is apparent from the left panel where even for the same initial PBH mass $m_{\rm in}$ (and hence the same $T_3$), the break occurs at different frequencies. Note that there appears another break in the spectrum in the middle, from power-law to flat, which can also be probed in the detectors. 

%Finally, recall that we are using only the $k=1$ mode for computing the GW spectra, and from Eq. \eqref{gwint} one can see that the spectra for the $k^{\rm th}$ mode is related to the fundamental $k=1$ mode through \cite{Gouttenoire:2019kij} 
%\begin{align}
%\Omega_{GW}^{(k)} (f)=k^{-\delta}\Omega_{GW}^{(1)} (f/k).
%\end{align}
%For the part of the spectra behaving as $\Omega_{GW}^{(1)} \propto f^0$, the total spectrum just gets re-scaled as  $\Omega_{GW} (f)= \zeta(\delta) \Omega_{GW}^{(1)} (f)$, where $\zeta(\delta) = \sum_k k^{-\delta}\simeq 3.6$ for cusps. On the other hand, for the GW spectra part varying as $\Omega_{GW}^{(1)} \propto f^{-1}$, summing over all modes changes this slope to $f^{-1/3}$ \cite{Gouttenoire:2019kij}. This effect of the higher k modes is illustrated in Fig. \ref{fig:highk}. Thus, we can conclude that including the effects of the higher k modes keeps the turning point frequencies unchanged, while leading to an increase in the overall amplitude of the GW spectra,  and hence still being in favour of being detected by the various GW experiments.

In the left panel of Fig. \ref{fig:MDM_TPF}, we show the contours in the $M_{\rm DM}-f_{\Delta}$ plane (grey-shaded) which satisfies the correct DM relic, along with the sensitivities of the GW experiments which can probe them. We find our numerical results to agree with our analytical estimates derived earlier up to a difference of around $\mathcal{O}(1)$. As expected, the contours move up when we increase the PBH mass $m_{\rm in}$. The upper bound on the DM mass appears since $N_{3}$ has to decay before BBN. Note that this bound changes with $m_{\rm in}$, since a heavier PBH evaporates at a lower temperature, closer to BBN. Decreasing DM mass leads to $\Omega_{\rm DM}h^2<0.12$, which gives us a lower bound on DM mass. In the right panel of Fig. \ref{fig:MDM_TPF}, for the same PBH masses we show the corresponding predictions at the higher frequencies, i.e. the $M_{\rm DM}-f'_{\Delta}$ plane. As we can see, for lighter PBH and heavier DM, $f'_{\Delta}$ falls outside the sensitivities. Finally, in Table \ref{BP}, we provide some benchmark values of DM mass $M_{\rm DM}$ and PBH mass $m_{\rm in}$ along with the GW experiments which can probe the corresponding spectral breaks.

%%%%%%%%%%%%%%%%%%%%%%%%%%%%%%%%%%%%%%%%%%%
\begin{figure}[htb!]
$$
\includegraphics[scale=0.5]{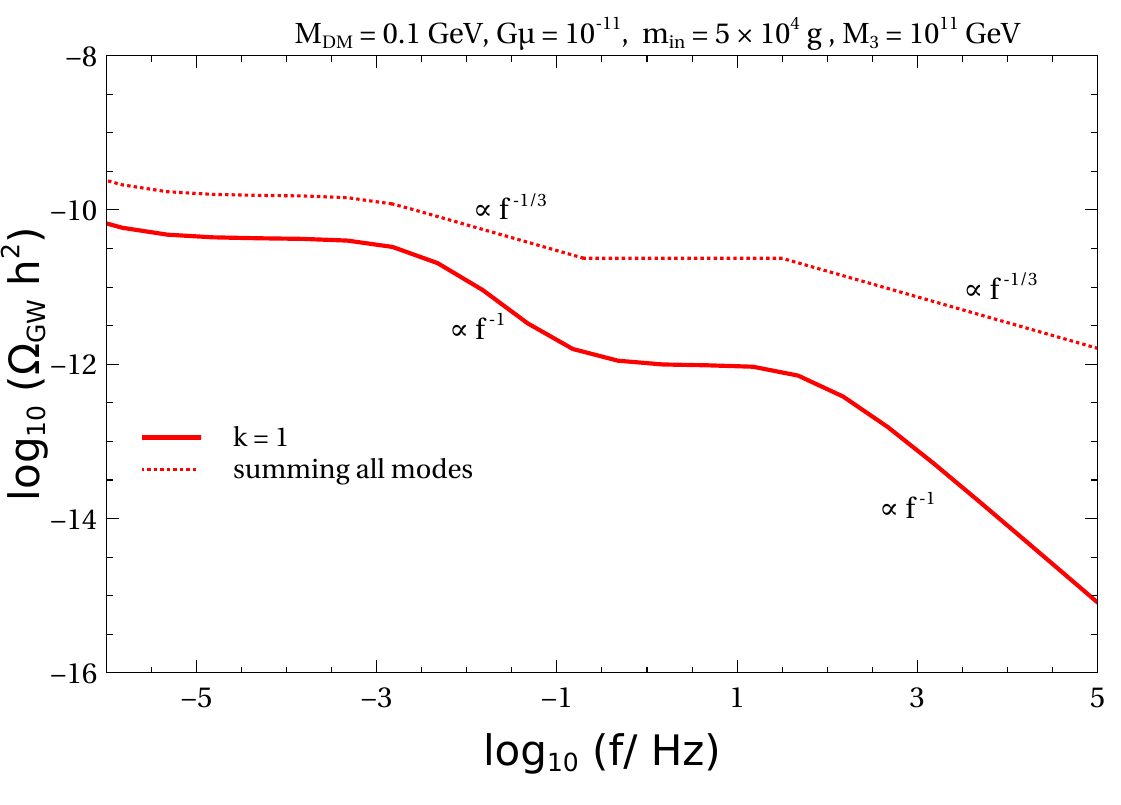}
$$
\caption{Effect of summing over the high k modes on the GW spectrum.}
\label{fig:highk}
\end{figure}
%%%%%%%%%%%%%%%%%%%%%%%%%%%%%%%%%%%%%%%%%%%%

Finally, recall that we are using only the $k=1$ mode for computing the GW spectra. Although summing upto high modes is computationally heavy and beyond the scope of this work, here we would like to mention the expected change and the consequences. From Eq. \eqref{gwint} one can see that the spectra for the $k^{\rm th}$ mode is related to the fundamental $k=1$ mode through \cite{Blasi:2020wpy, Gouttenoire:2019kij} 
\begin{align}
\Omega_{GW}^{(k)} (f)\simeq k^{-\delta}\Omega_{GW}^{(1)} (f/k).
\end{align}
For the part of the spectra behaving as $\Omega_{GW}^{(1)} \propto f^0$, the total spectrum just gets re-scaled as  $\Omega_{GW} (f)\simeq \zeta(\delta) \Omega_{GW}^{(1)} (f)$, where $\zeta(\delta) = \sum_k k^{-\delta}\simeq 3.6$ for cusps. On the other hand, for the GW spectra part varying as $\Omega_{GW}^{(1)} \propto f^{-1}$, summing over all modes changes this slope to $f^{-1/3}$ \cite{Gouttenoire:2019kij}. This effect of the higher k modes is illustrated in Fig. \ref{fig:highk}. Thus, we can conclude that including the effects of the higher k modes keeps the turning point frequencies nearly unchanged, while leading to an increase in the overall amplitude of the GW spectra,  and hence still being in favor of being detected by the various GW experiments.

\begin{table}[htb!]
\centering
\begin{tabular}{|c | c| c| c | c| c |}
\hline
BP   & $m_\text{in}$(g) & $M_\text{DM}$(GeV) & $1^\text{st}~\text{TPF}$ &$2^\text{nd}~\text{TPF}$ &$3^\text{rd}~\text{TPF}$\\ \hline \hline
BP1 & $3\times 10^3$ & 0.01 & ET & NONE & NONE        \\ \hline
BP2 & $1.2\times 10^4$ & 5 & LISA, DECIGO, BBO & CE & NONE        \\ \hline
BP3 & $5\times 10^4$ & 0.1 & LISA & DECIGO, BBO & CE        \\ \hline
BP4 & $10^5$ & 0.3 & LISA, $\mu$ARES & DECIGO, BBO & NONE \\ \hline
BP5 & $5\times 10^5$ & 0.15 & $\mu$ARES & LISA & DECIGO, BBO        \\ \hline
	\end{tabular}
	\caption{Some benchmark points (BP) of $M_{\rm DM}$ and $m_{\rm in}$ along with the GW experiments which can probe the $1^{\rm st}$ TPF($f_{\Delta}$) from flat to power-law, $2^{\rm nd}$ TPF from power-law to flat and $3^{\rm rd}$ TPF($f'_{\Delta}$) from flat to power-law.}
	\label{BP}
\end{table}
\section{Gravitational Waves from PBH}
\label{sec4}
Apart from the GW background generated from Cosmic Strings, PBH themselves might be involved in the production of GW in several ways. The evaporation of PBH can produce gravitons which might constitute an  ultra-high frequency  GW spectrum \cite{Anantua:2008am}. PBH can also form mergers which can lead to GW emission \cite{Zagorac:2019ekv}. Next, the scalar perturbations leading to the formation of PBH can induce GWs at second-order \cite{Saito:2008jc}, which can be enhanced during PBH evaporation \cite{Inomata:2020lmk}. Finally, the inhomogeneity in the distribution of PBH may also induce GW at second order, as recently studied in  Ref. \cite{Papanikolaou:2020qtd, Domenech:2020ssp, Domenech:2021wkk}. We concentrate on this last possibility since such GW generated are independent of the formation history of PBH and especially, as we will see,  for the PBH mass range we are considering, it  can lead to GW signals within the sensitivity of near future GW detectors.

The distribution of PBH after they are formed are random and follow Poissonian statistics \cite{Papanikolaou:2020qtd}. These inhomogeneities induce GW at second order, when PBH dominate the energy density of the Universe, and are enhanced during PBH evaporation \cite{Domenech:2020ssp}. The dominant contribution to the GW amplitude observed today can be written  as \cite{Domenech:2020ssp, Borah:2022iym, Barman:2022pdo}
\begin{equation}
    \ogw(t_0,f)\simeq \ogw^{\rm peak}\left(\frac{f}{f^{\rm peak}}\right)^{11/3}\Theta
\left(f^{\rm peak}-f\right),\label{eqn:omgwpbh}
\end{equation}
where
\begin{equation}
    \ogw^{\rm peak}\simeq \frac{2\times 10^{-6}}{S^{4/3}} \left(\frac{\beta}{10^{-8}}\right)^{16/3}\left(\frac{m_{\text{in}}}{10^7 \rm g}\right)^{34/9}.\label{eqn:omgpeakpbh}
\end{equation}
Here the factor $S$ (c.f. Eq. \eqref{eq:dilutionF}) accounts for any entropy injection into the SM bath after the PBH evanescence, which in our case is provided by the decay of the long-lived diluter field. The GW spectrum has an ultraviolet cutoff at frequencies corresponding to comoving scales representing the mean separation between PBH \cite{Papanikolaou:2020qtd, Domenech:2020ssp}, which is given by 
\begin{equation}
    f^{\rm peak}\simeq \frac{1.7\times 10^3\,{\rm Hz}}{S^{1/3}}\,\left(\frac{m_{\text{in}}}{10^4 \rm g}\right)^{-5/6}.\label{eqn:fpkpbh}
\end{equation}

\begin{figure}[htb!]
$$
\includegraphics[scale=0.6]{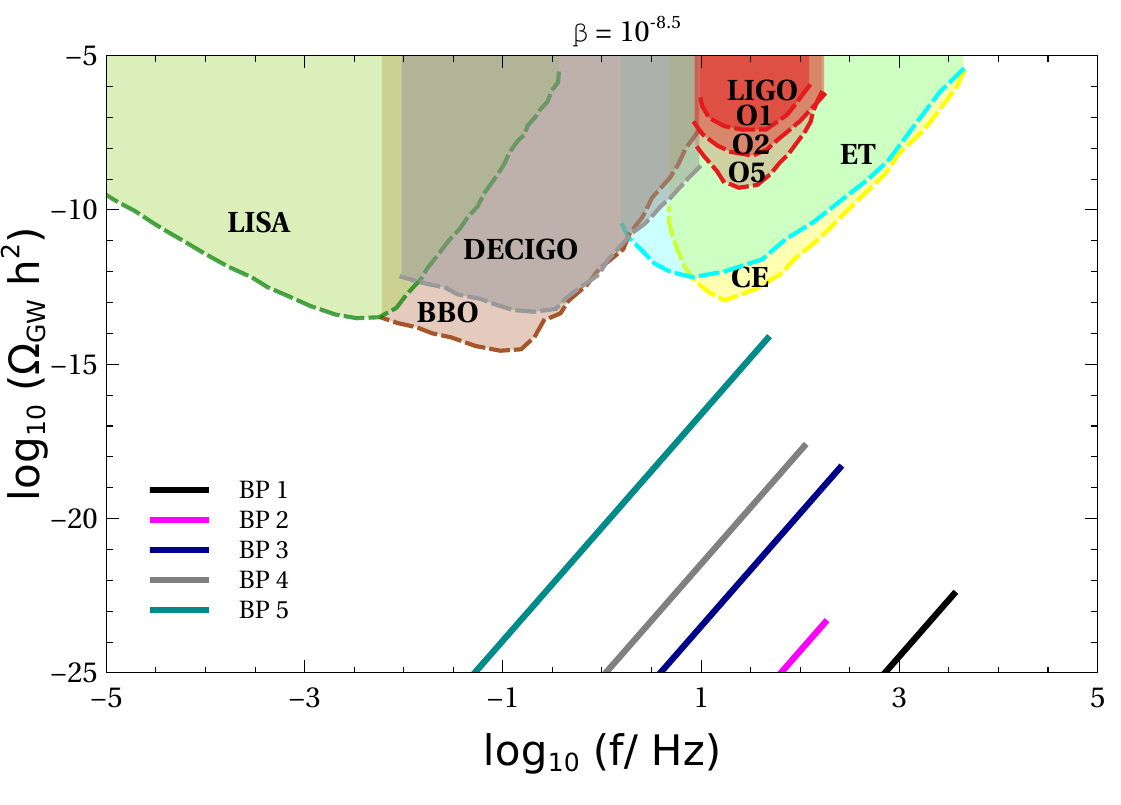}
$$
\caption{GW spectra from PBH density fluctuations for the benchmark values of Table \ref{BP}, considering $\beta=10^{-8.5}$.}
\label{fig:pbh_igw}
\end{figure}

Note that the peak of the GW amplitude (Eq. \eqref{eqn:omgpeakpbh}) increases with $\beta$, which has been a free parameter in our analysis so far.  In Fig. \ref{fig:pbh_igw}, we show the GW spectra generated from PBH density fluctuations for the benchmark points given in Table \ref{BP}, considering $\beta=10^{-8.5}$. We ensure that $\beta > \beta_c$ holds for all the BPs, which is required for our analysis to be valid. Moreover, we also ensure that $\beta<\beta_{\rm max}\simeq 1.1 \times 10^{-6} \left(\frac{m_{\rm in}}{10^{4}g}\right)^{-17/24}$, where $\beta_{\rm max}$ is the upper bound on $\beta$ which comes from the contribution of this GW spectrum to extra relativistic degrees of freedom during BBN \cite{Domenech:2020ssp, Domenech:2021wkk}. It is pertinent to note that so far our analysis of dark matter and GW spectrum from cosmic strings has been independent of the parameter $\beta$, provided it satisfies $\beta > \beta_c$. Hence, in the study of GW from PBH density fluctuations, $\beta$ appears to be a free parameter, which we consider to be small such that the GW peak amplitude which  varies as $\beta^{16/3}$ (cf. Eq. \eqref{eqn:omgpeakpbh}) remains subdominant compared to that from cosmic strings.  The motivation for particularly focusing on the GW spectrum from cosmic strings is that it has distinct signatures of multiple matter dominated eras because of the presence of multiple turning point frequencies, which is usually absent for GW from other sources.  Also, as discussed briefly in section \ref{sec2}, it is indeed possible to generate PBH with such initial energy fraction and mass within some realistic scenarios.

%%%%%%%%%%%%%%%%%%%%%%%%%%%%%%%%%%%%%%%%%%%%%%%%%%%%%%%%%%%%%%%%%%%%%%%%%%%%%%%%%%%%%%%%%%%%%%%%%%%%%%%%%%%%%%%%%%%%%%%%%%%%%%%%%%%%%%%%%%
%\section{Neutrino Mass and Leptogenesis}

\section{Conclusion}
\label{sec5}
We have studied the possibility of probing a high-scale seesaw origin of neutrino mass and PBH-generated dark matter scenario simultaneously by future observation of stochastic gravitational wave background with multiple tilts or spectral breaks. A high scale $U(1)$ gauge symmetry breaking not only generates the seesaw scale dynamically but also leads to the formation of cosmic strings responsible for generating stochastic GW later. Considering purely gravitational dark matter originating from an intermediate primordial black hole dominated epoch, we also incorporate the presence of an additional diluter to bring DM overproduction from PBH under control. The required relic abundance of DM around the MeV-TeV ballpark, which we consider in this work, forces the diluter to be sufficiently long-lived leading to a second early matter-domination epoch in addition to the PBH-dominated one. While cosmic strings typically lead to a scale-invariant GW spectrum, the presence of two different early matter domination in our setup leads to multiple spectral breaks. After solving the relevant Boltzmann equations numerically for DM relic, we make an estimate of the spectral break frequencies both analytically as well as numerically and find interesting correlations with DM mass. As summarised in Fig. \ref{fig:MDM_TPF} and Table \ref{BP}, the spectral break frequencies can be probed at multiple future experiments depending upon the DM mass. Clearly, the parameter space corresponding to a wide range of DM masses remains within experimental reach, although the three spectral breaks simultaneously remain within the experimental sensitivities only when the DM mass is restricted to be in the sub-GeV ballpark. While high-scale seesaw and gravitational DM have no scope of the direct experimental probe, our setup provides an interesting near-future probe of such scenarios via multiple spectral breaks in the gravitational wave spectrum generated by cosmic strings. We also calculate the GW spectrum generated by PBH density fluctuations. While such GW spectrum can, in principle, have large amplitude in the frequency range of our interest, our choice of the parameter $\beta$ keeps this contribution sub-dominant compared to the one generated by cosmic strings. We leave detailed studies of such combined GW spectrum including a UV complete model explaining the origin of PBH to future works.

%%%%%%%%%%%%%%%%
\section*{Acknowledgements}
%%%%%%%%%%%%%%%%
The work of RR was supported by the National Research Foundation of Korea (NRF) grant funded by the Korean government (NRF-2020R1C1C1012452).
%%%%%%%%%%%%%%%%%%%%%%%%%%%%%%%%%%%%%%%%%%%%%%%%%%%%%%%%%%%%%%%%%%%%%%%%%%%%%%%%%%%%%%%%%%%%%%%%%%%%%%%%%%%%%%%%%%%%%%%%%%%%%%%%%%%%%%%%%%%%%
\appendix

%%%%%%%%%%%%%%%%%%%%%%%%%%%%%%%%%

\section{A UV complete seesaw model}
\label{sec:model}
Here we give one possible UV completion of the seesaw model discussed earlier. For a demonstrative purpose, here we consider an anomaly free $U(1)_{L_\mu-L_\tau}$ extension of the SM gauge symmetry where $L_\mu$ and $L_{\tau}$ represent the muon and tau lepton numbers respectively \cite{He:1990pn, He:1991qd, Foot:1990mn}. The SM particle content is extended with three additional right-handed neutrinos ($N^e_R,N^\mu_R,N^{\tau}_R$) and a complex scalar ($S$) that is a singlet under the SM gauge symmetry but carries $1$ unit of $U(1)_{L_\mu-L_\tau}$ charge. As the symmetry suggests, the muon and tau take $1$ and $-1$ units of charge under the  $U(1)_{L_\mu-L_\tau}$ respectively. The newly introduced RHNs are singlets under the SM gauge symmetry, while two carry 1 and $-1$ unit of $U(1)_{L_\mu-L_\tau}$ charges and the third remains uncharged. It is worth noting that there is nothing special about the choice of this gauge symmetry to realise our setup. One could, in principle, consider a hidden $U(1)$ gauge symmetry too under which right handed neutrinos are charged such that the scale of $U(1)$ breaking is the same as the seesaw scale. However, this will require the presence of additional chiral fermions to keep the model anomaly free and additional scalar doublet(s) to allow Dirac Yukawa coupling. Therefore, the choice of a $U(1)$ gauge symmetry under which SM leptons are charged remains a minimal choice. Of course, $U(1)_{L_\mu-L_\tau}$ is not the only symmetry of this type. We can realise our setup within similar gauge symmetries like $U(1)_{B-L}$ too. However, since all SM leptons are charged under $U(1)_{B-L}$ and we require the diluter to be chargeless under the $U(1)$ symmetry as mentioned earlier, here the diluter can not be a right handed neutrino. In such a scenario, one can simply have a scalar diluter which can decay into SM Higgs at late epochs.

Next, we write all the possible interactions the different particles can have in the present setup. The kinetic terms for the additional fields read as,
\begin{align}
\mathcal{L^{\rm KE}}&=\frac{i}{2}\sum_{\alpha=e,\mu,\tau}N_\alpha\gamma^\delta D_\delta N_\alpha+(D^\delta S)^\dagger(D_\delta S)&
\label{kinetic_term}
\end{align} 

where $D_{\delta}=\partial_\delta+ig_{\mu\tau}Q_{\mu\tau}(Z_{\mu\tau})_\delta$ with $Q_{\mu\tau}$ representing  the charge and $Z_{\mu\tau}$ being the gauge boson of $U(1)_{L_\mu-L_\tau}$ symmetry. The Lagrangian involving the Yukawa interactions and masses of the additional fermions involved can be written as,
\begin{align}
\mathcal{L}=&-\frac{1}{2}h_{e\mu}(\bar{N}^c_e N_\mu+\bar{N}^c_\mu N_e)S^\dagger-\frac{1}{2}h_{e\tau}(\bar{N}^c_e N_\tau+\bar{N}^c_\tau N_e)S-\sum_{\alpha=e,\mu,\tau}Y_\alpha\bar{L}_\alpha \tilde{H}N_\alpha \nonumber\\
&-\frac{1}{2}M_{ee}\bar{N}^c_e N_e-\frac{1}{2}M_{\mu\tau}(\bar{N}^c_\mu N_\tau+\bar{N}^c_\tau N_\mu)+h.c.
\label{yukawa_term}
\end{align}

The most general scalar potential involving the different scalars can be expressed as,
\begin{align}
V(H,S) =& -\mu^2_H H^\dagger H -\mu^2_S S^\dagger S
+ \lambda_H (H^\dagger H)^2 + \lambda_S (S^\dagger S)^2
 + \lambda_{HS} (H^\dagger H)(S^\dagger S).
 \label{potential}
\end{align}

The scalar $S$ breaks the $U(1)_{L_\mu-L_\tau}$ symmetry once its CP even component develops a non-zero VEV $v_{\mu\tau}$.  This breaking also results in an additional non-zero mixing between the RHNs as can be seen from Eq.~\eqref{yukawa_term}. Once the Electroweak Symmetry is broken, the Higgs doublet ($H$) also develops a non-zero vev $v = 246$ GeV. As a result of electroweak symmetry breaking, the different scalars will mix with each other. We will not go into the details of the scalar mixing as it remains unimportant for our analysis and refer the reader to \cite{Asai:2017ryy} for the details. In fact, due to the restrictive nature of Dirac neutrino and RHN mass matrices, fitting light neutrino data requires another singlet scalar, as pointed out in \cite{Asai:2017ryy, Borah:2021mri}. The details of these extensions do not affect our analysis and hence we skip them here.

Finally, we consider scalar singlet dark matter $\phi$ to have only gravitational interactions and hence it has a bare mass term (denoted by $m_\text{DM}$) only in the Lagrangian. Another heavy RHN, similar to $N^e$ with vanishing $U(1)_{L_\mu-L_\tau}$ charge but tiny couplings to leptons is considered to be negligible. This diluter RHN, to be denoted as $N_3$, also couples more feebly to other RHNs so that $N_3$ dominantly decays into $L_e, H$ at late epochs. While $N_e$ could, in principle, play the role of diluter as well, it will lead to difficulties in generating correct light neutrino data due to restricted mass matrix structures. Due to tiny couplings of $N_3$ with other particles in the thermal bath, it is produced only non-thermally from the evaporation of the PBH. The consequence of this slow decay is a $N_3$ dominated epoch after the evaporation of PBH.

\section{PBH fact-sheet}
\label{sec:pbh}
%%%%%%%%%%%%%%%%%%%%%%%%%%%%%%%%%

The mass of the black hole from gravitational collapse is typically close to the value enclosed by the post-inflation particle horizon and is given by~\cite{Fujita:2014hha,Masina:2020xhk}
\begin{equation}
m_{\rm in}=\frac{4}{3}\,\pi\,\gamma\,\Bigl(\frac{1}{\mathcal{H}\left(T_\text{in}\right)}\Bigr)^3\,\rho_\text{rad}\left(T_\text{in}\right)\,
\label{eq:pbh-mass}
\end{equation}
\noindent with 
\begin{equation}
\rho_\text{rad}\left(T_\text{in}\right)=\frac{3}{8\,\pi}\,\mathcal{H}\left(T_\text{in}\right)^2\,M_\text{pl}^2\,
\end{equation}
where $\mathcal{H}$ is the Hubble parameter. As mentioned earlier,  PBHs are produced during the radiation dominated epoch, when the SM plasma has a temperature $T=T_\text{in}$ which is given by
\begin{equation}
T_\text{in}=\Biggl(\frac{45\,\gamma^2}{16\,\pi^3\,g_\star\left(T_\text{in}\right)}\Biggr)^{1/4}\,\sqrt{\frac{M_\text{pl}}{m_{\rm in}}}\,M_\text{pl}\,.
\label{eq:pbh-in}
\end{equation}
Once formed, a PBH evaporates efficiently into particles lighter than its instantaneous temperature $T_\text{BH}$ defined as \cite{Hawking:1975vcx}
\begin{equation}
T_{\rm BH}=\frac{1}{8\pi\,G\, m_{\rm BH}}\approx 1.06~\left(\frac{10^{13}\; {\rm g}}{m_{\rm BH}}\right)~{\rm GeV}\,,
\end{equation}
\noindent with $G$ being the universal gravitational constant. Note that $m_{BH}$ changes with time whereas $m_{\rm in}$ denotes the initial PBH mass at the time of formation. The mass loss rate in this scenario is parametrized as  \cite{MacGibbon:1991tj}
\begin{equation}
\frac{dm_\text{BH}(t)}{dt}=-\frac{\mathcal{G}\,g_\star\left(T_\text{BH}\right)}{30720\,\pi}\,\frac{M_\text{pl}^4}{m_\text{BH}(t)^2}\,.
\label{eq:pbh-dmdt}
\end{equation}
Under the assumption that $g_\star$ does not change during the PBH evolution, integrating Eq.~\eqref{eq:pbh-dmdt}, we end up with the PBH mass evolution equation as
\begin{equation}
m_\text{BH}(t)=m_{\rm in}\Bigl(1-\frac{t-t_\text{in}}{\tau}\Bigr)^{1/3}\,,
\end{equation}
with 
\begin{equation}
\tau = \frac{10240\,\pi\,m_{\rm in}^3}{\mathcal{G}\,g_\star(T_\text{BH})\,M_\text{pl}^4}\,,\label{eq:PBHlt}
\end{equation}
as the PBH lifetime. One can calculate the evaporation temperature by considering $H(T_\text{ev})\sim\frac{1}{\tau^2}\sim\rho_\text{rad}(T_\text{ev})$,
\begin{equation}
T_\text{ev}\equiv\Bigl(\frac{45\,M_\text{pl}^2}{16\,\pi^3\,g_\star\left(T_\text{ev}\right)\,\tau^2}\Bigr)^{1/4}\,.
\label{eq:pbh-Tev}
\end{equation}
However, if the PBH component dominates at some point the total energy density of the universe, the SM temperature just after the complete evaporation of PBHs is: $\overline{T}_\text{ev}=2/\sqrt{3}\,T_\text{ev}$~\cite{Bernal:2020bjf}.  For PBH to dominate the energy density of the Universe before they evaporate, one should have the following condition on $\beta$ \cite{Masina:2020xhk}
\begin{align}
    \beta > \beta_c = \gamma^{-1/2}\,\sqrt{\frac{\mathcal{G}\,g_{\star,H}(T_\text{BH})}{10240\,\pi}}\,\frac{M_\text{pl}}{m_\text{in}}\,. \label{eqn:betac1}
\end{align}

%%%%%%%%%%%%%%%%%%%

%\bibliographystyle{JHEP}
%\bibstyle{apsrev}
%\bibliography{ref, ref1}
\providecommand{\href}[2]{#2}\begingroup\raggedright\endgroup

\end{document}